%% file: piston-0.tex
\documentclass[article, preprint]{siamart0516}
\usepackage{lipsum}
\usepackage{amsfonts}
\usepackage{graphicx}
\usepackage{epstopdf}
\usepackage{subfigure}
\usepackage{bm}
\ifpdf
  \DeclareGraphicsExtensions{.eps,.pdf,.png,.jpg}
\else
  \DeclareGraphicsExtensions{.eps}
\fi

 \newtheorem{thm}{Theorem}[section]

 \theoremstyle{definition}
 
 \theoremstyle{remark}
 \newtheorem{rem}[thm]{Remark}

\newcommand{\TheTitle}{{\large Accelerated Piston Problem and High Order Moving Boundary Tracking Method for Compressible Fluid Flows}}
\newcommand{\TheAuthors}{Zhifang Du and Jiequan Li}

\headers{accelerated piston problem and high order MBT}{\TheAuthors}

\title{{\TheTitle}
}

\author{
  Zhifang Du\thanks{Institute of Applied Physics and Computational Mathematics, 100088, Beijing, P. R. China; Email: du@mail.bnu.edu.cn. This work of Zhifang Du is supported by China Postdoctoral Science Foundation (2018M641271).}
   \and
  Jiequan Li\thanks{Laboratory of Computational Physics,  Institute of Applied Physics and Computational Mathematics, Beijing, P. R. China: Center for Applied Physics and Computational Mathematics, Peking University; Email: li\_jiequan@iapcm.ac.cn. Jiequan LI is supported by NSFC (nos. 11771054, 91852207) and Foundation of LCP. Also he thanks the hospitality from Professor Kun Xu during his stay in Hongkong University of Science and Technology. }
}

\usepackage{amsopn}

\include{abbr-notation}

\ifpdf
\hypersetup{
  pdftitle={\TheTitle},
  pdfauthor={\TheAuthors}
}
\fi

\begin{document}

\maketitle 

% REQUIRED
\begin{abstract} %Interactions between compressible fluids and solid bodies are considered in the present paper.
Reliable tracking of moving boundaries  is important for the simulation of compressible fluid flows and there are a lot of contributions in literature.  We recognize from the classical piston problem, a typical moving boundary problem in gas dynamics, that  the acceleration is a key element in the description of the motion and it should be incorporated into the design of a moving boundary tracking (MBT) method. Technically, the resolution of the accelerated piston problem boils down to a one-sided generalized Riemann problem (GRP) solver, which is taken as the building block to construct schemes with the high order accuracy both in space and time.   In this paper we take this into account, together with the cell-merging approach,  to propose a new family of high order accurate moving boundary tracking methods and verify its performance through one- and two-dimensional test problems, along with accuracy analysis. 
 
\end{abstract}

% REQUIRED
\begin{keywords}
 Compressible fluid flows, accelerated piston problem,  moving boundary tracking method,  one-sided GRP solver, cell-merging criterion.
\end{keywords}

\begin{AMS}
35L50, 35L65, 65M08, 76M12, 76N15
\end{AMS}
% REQUIRED
%\begin{AMS}
%  MSC
%\end{AMS}

\section{Introduction}\label{sec:intro}

Moving boundary problems are ubiquitous for engineering applications and particularly   the tracking of moving boundaries is an essential technique determining the quality of underlying simulations for compressible fluid flows.  There  are  a lot of studies  in this context, e.g., the front tracking method \cite{Glimm-1999}, the moving boundary tracking (MBT) method \cite{mbt-falc,Shyue-08}, the immersed boundary method \cite{Peskin-2002}, level set methods \cite{Osher-2018}, volume of fluid methods \cite{VOF-2001}, moment of fluid methods \cite{MOF-2008}, adaptive mesh refinement methods \cite{AMR-1989} and many others. In this paper, we will follow the moving boundary tracking method in the finite volume framework, with the new recognition of the key role  of the acceleration in the description of  moving boundaries, to develop a high order accurate version. Our study is motivated by a basic moving boundary problem, the accelerated piston problem in gas dynamics \cite{Courant-Friedrichs}.

Assume that a uniform piston with a thickness $L$,   moves in a tube filled with gas. Its motion is described by two elements, the velocity and the acceleration \cite{Takeno-1995}, 
\begin{equation}\label{eq:dynam-piston}
\bga{l}
\dfr{dx_c(t)}{dt}=u_c(t),\\[2.5mm]
\dfr{du_c(t)}{dt}=-\mathcal{S}_c\dfr{p(x_c(t)+L/2+0,t)-p(x_c(t)-L/2-0,t)}{\mathcal{M}_c},
\eda
\end{equation}
where $x_c(t)$, $u_c(t)$ and $\mathcal{M}_c$ are the location, the velocity and the mass of the piston, $\mathcal{S}_c$ is the sectional area of the tube and $p(x_c(t)\pm L/2\pm 0,t)$ are pressures exerted on the two faces of the piston. 
The two equations in \eqref{eq:dynam-piston} are the Newtonian first and second laws for fluid flows, respectively,  %And the second one is the counterpart of the momentum equation in Euler equations for a solid bo
and they are coupled with the Euler equations,  leading to a coupled dynamical system \cite{FSI-SIAM}.  This might be the simplest moving (free) boundary problem in the context of compressible flows.    It is natural 
to use the pair $(u_c(t), du_c(t)/dt)$ for the tracking of this piston, more or less like the symplectic algorithm \cite{Symplectic-2010},
\begin{equation}
x_c(t_{n+1}) =x_c(t_n) +\De t u_c(t_n) + \frac{\De t^2}{2} \dfr{du_c(t_n)}{dt} +\mathcal{O}(\De t^3),
\label{eq:b-t} 
\end{equation} 
where $\De t =t_{n+1}-t_n$ is the time increment. 
Otherwise, one would have to use the Runge-Kutta type time stepping to advance the trajectory of the piston, for which only the velocity $u_c(t)$ is adopted. As the piston is resolved together with the evolution of the fluid flow in the whole flow region, the problem boils down to the one-sided generalized Riemann problem (GRP), the so-called initial boundary value problem (IBVP) with the piston as a free boundary.  Numerically, when  the finite volume framework is adopted, it is necessary to   develop  a one-sided GRP solver in order to construct numerical fluxes on the piston surface, and track the moving boundary (piston).  The current study just establishes the interrelation  between the resolution of the accelerated piston problem and the MBT method, leading to   a high order version of MBT.

 The moving boundary tracking method was originated in \cite{mbt-falc}, and  its advantage was clearly stated there. Since the uniformity of flow variables is assumed in boundary cells and the  one-sided Riemann solver is adopted to compute the corresponding numerical fluxes, the MBT algorithm developed there has at most the first order accuracy  in boundary cells \cite[page 91]{mbt-falc}.  The present contribution uses the direct Eulerian GRP solver \cite{Li-1}, which enables us to develop a one-sided GRP solver suited for the MBT methodology to achieve the high order accuracy. The boundary can be tracked with high order accuracy simultaneously and no extra technology is needed.  Besides, since the present GRP solver has been updated for multi-dimensional computations \cite{Du-Li-1},  the dimensional operator-splitting algorithm in \cite{mbt-falc} is not necessary. 
  As far as complex geometries are concerned, the present paper adopts  the cell-merging approach in  \cite{cut-cell, quirk} to deal with the ``small-cell'' problem. Small cut cells are absorbed by their neighbors to form larger control volumes and fluid states are evolved over  time-varying control volumes so that  the resulting  MBT scheme is consistent with the integrated form of conservation laws. This is different from the body-fitted or  adaptive grid schemes  \cite{mbt-adaptive,Olim-1993,overlap-NASA,overlap-Fedkiw}. The MBT is also applied in other frameworks, e.g., the wave propagation algorithm \cite{Shyue-08}.   We particularly refer to the inverse Lax-Wendroff (ILW) method \cite{mbt-ilw}, which  interpolates the ghost fluid state with high-order spatial accuracy by taking the boundary acceleration into consideration. The temporal accuracy is obtained by the popular Runge-Kutta discretization.

This paper is organized as follows. 
Basic facts about the rigid body motion and the  MBT method are given in Section \ref{sec:preliminary}.
 One- and two-dimensional one-sided GRP solvers are developed in Section \ref{sec:solver}. 
The moving boundary tracking scheme is proposed in Section \ref{sec:mbt-fv}. The numerical accuracy of the newly developed MBT method is analyzed in Section \ref{sec:accuracy_analysis}.
Numerical experiments involving fluid-surface interactions are carried out  in Section \ref{sec:numer} for the performance. Some discussions are made in Section \ref{sec:discussion}.\\

\section{ Accelerated piston problem and moving boundary tracking method} \label{sec:preliminary}
This section  describes the piston problem, which motivates the new high-order MBT scheme that we are going to propose. The key observation is that the acceleration of a boundary should be  taken as a natural and necessary element to describe its motion. Let's  describe the rigid body motion in one and two space dimensions.

\subsection{Rigid body motion and accelerated piston problem}
\begin{figure}
 \centering
\includegraphics[width=.35\linewidth]{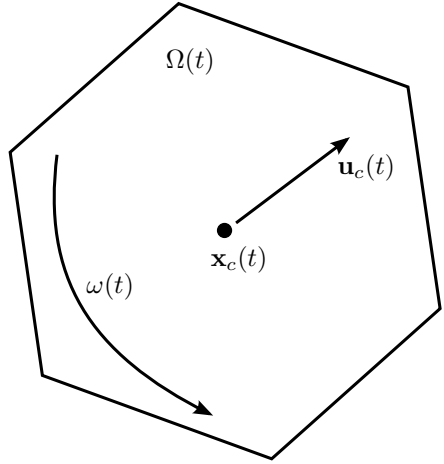}
\put(-88, 75){$\bx_c(t)$}
\put(-40, 110){$\bu_c(t)$}
\put(-135, 65){${\om}(t)$}
\put(-105, 150){$\Om(t)$}
\caption[small]{ The motion of a  rigid body $\Om(t)$ in the compressible fluid is described  with  its  barycenter $\bx_c(t)$, a translational  $\bu_c(t)$ and  a rotational velocity $\om(t)$.   }
\label{Fig:rigid}
\end{figure}

Think of a rigid body, represented by the polygon $\Om(t)$ with its barycenter locating at $\bx_c(t)$, moving in the compressible fluid with a translational velocity $\bu_c(t)$ and a rotational velocity $\om(t)$, as shown in Figure \ref{Fig:rigid}. The rotational velocity $\om$ is regarded as a pseudovector, which is positive if $\Om$ rotates counter-clockwise. The flow field around this body is described by the compressible Euler equations (only two-dimensional in the present paper, and similarly for three dimensional cases), 
\begin{equation}
\begin{array}{l}
\dfr{\pt \bW}{\pt t}  +\dfr{\pt\bF(\bW)}{\pt x}  +\dfr{\pt\bG(\bW)} {\pt y}=0,\\[3mm]
\bW=(\rho,\rho u,\rho v, \rho E)^\top, \\[3mm]
 \bF(\bW) =(\rho u,\rho u^2+p,\rho uv, u(\rho E+p))^\top,  \\[3mm]
  \bG(\bW) =(\rho v,\rho uv, \rho v^2+p, v(\rho E+p))^\top,
\end{array} 
\label{eq:euler} 
\end{equation} 
where $\bu=(u,v)$ is the velocity in the $\bx=(x,y)$-coordinates, $\rho$ is the density,  $p$ is the pressure, $E=\frac 12 (u^2+v^2) +e$ is the total energy. The internal energy $e$ is determined by the equation of state (EOS) $e=e(\rho,p)$. Thermodynamical quantities satisfy the Gibbs relation,
\begin{equation}\label{eq:Gibbs}
Tds=de +pd\tau,
\end{equation}
where $\tau=1/\rho$ and $s$ is the entropy. For polytropic gases,  $p=(\gamma-1)\rho e$ and $\gamma>1$ is the specific heat ratio.

The  boundary of the rigid body $\Om(t)$ is  denoted by $\pt\Om(t)$ with the unit normal vector $\bn$  pointing from the fluid to the rigid body.
% by $\bx_c(t)$ the barycenter,  by $\bu_c$ the velocity of the body,  and by $\mathcal{M}_c$ the mass of $\Om(t)$.  
Then the motion of the body  obeys 
\begin{equation}\label{eq:2d-translational}
\dfr{d\bx_c(t)  }{dt}=\bu_c(t),  \ \  \dfr{d\bu_c(t)}{dt}=\dfr{1}{\mathcal{M}_c} \int_{\pt\Om(t)}p\ \bn \ dl,
\end{equation}
 and its rotational motion is described using the equations
\begin{equation}\label{eq:2d-rotational}
\dfr{d\theta}{dt}=\om(t), \ \ \dfr{d \om}{dt}=\dfr{1}{\mathcal{A}_c}\int_{\pt\Om(t)}p\  (\bx-\bx_c(t)) \times \bn  \ dl,
\end{equation}
where $\mathcal{M}_c$ is the mass of the solid body, $\theta(t)$ and $\om(t)$ are the relative angle and the angular velocity of $\Om(t)$ with respect to its barycenter $\bx_c(t)$, and $\mathcal{A}_c$ is the inertia of the solid body. For $\forall\bX_b\in\pt\Om(0)$, denote by $\bx_b(t;\bX_b)$ its trajectory from $\bX_b$ for $t>0$.
The instantaneous translational motion of any point $\bx_b(t;\bX_b)\in\pt\Om(t)$ is described by
\begin{equation}\label{eq:motion}
\begin{array}{l}
\dfr{d\bx_b(t;\bX_b)}{dt} =\bu_c(t)+\om(t)\ \br^\perp_b(t;\bX_b), \ \ \ \ \bx_b(0;\bX_b)=\bX_b,\\[3mm]
\dfr{d\bu_b(t;\bX_b)}{dt} = \dfr{d\bu_c}{dt} + \dfr{d\om(t)}{dt}\ \br^\perp_b(t;\bX_b)+\om(t)\Big[(\om(t)\ \br^\perp_b(t;\bX_b))\Big]^\perp, 
\end{array}
\end{equation}
where $\br_b(t;\bX_b)=\bx_b(t;\bX_b)-\bx_c(t)$, $\bu_c(0)$ and $\om(0)$ are prescribed,  $\bu_b$ is the velocity of the boundary point $\bx_b(t;\bX_b)\in \pt\Om(t)$ . %The results stated in \eqref{eq:pointwise-motion} will be useful in later calculations.
We use the notation $\ba^\perp=(-a_y,a_x)^\top$ for any vector $\ba=(a_x,a_y)^\top$.
Assume for the time being that the motion is not affected by surroundings  in order to avoid the complications caused by the interaction with other boundaries.  In order to describe the interaction between the rigid body and the compressible fluid, a proper boundary condition along the moving boundary $\pt\Om(t)$ is required, 
\begin{equation}\label{eq:bc-2d}
\bu(\bx_b(t;\bX_b),t)\cdot\bn(t;\bX_b)=\bu_b(t;\bX_b)\cdot\bn(t;\bX_b), \ \text{for } \forall\bx_b(t;\bX_b)\in\pt\Om(t),
\end{equation}
where $\bn(t;\bX_b)$ is the unit normal vector of $\pt\Om(t)$ at $\bx_b(t;\bX_b)$, pointing from the fluid into the rigid body.

When the problem reduces to one dimensional cases, the rigid body motion in the flow field corresponds to the  {\it free piston problem}  \cite{Courant-Friedrichs, Takeno-1995}, as described in Introduction.    As far as the motion of the piston is concerned, we ignore the discussion on both ends of the cylinder and assume that this piston moves over the whole line.   The thickness $L$ can be also assumed to be infinitely small.
\iffalse
The thickness $L$ can be also assumed to be infinitely small. Then \eqref{eq:plug} becomes, 
\begin{equation}
\label{eq:piston}
\bga{l}
\dfr{d x_c(t)}{dt}=u_c(t),\\[2.5mm]
\dfr{d u_c(t)}{dt}=\dfr{p(x_c(t)+0,t)-p(x_c(t)-0,t)}{\mathcal{M}_c}.
\eda
\end{equation}
\fi
Then $p(x_c(t)+0,t)$ and $p(x_c(t)-0,t)$ are obtained by simultaneously solving the following two  (left and right) free boundary problems,
\begin{equation}\label{eq:piston-l}
%\left\{
\begin{array}{ll}
\dfr{\pt \bW}{\pt t} +\dfr{\pt \bF(\bW) }{\pt x}=0,  &  x_c(t)<x<\iy, \ \ t>0,\\[3mm]
\bW(x,0)=\bW_+(x), & x_c(0)<x<\iy,\\[3mm]
u(x_c(t),t)=u_c(t), & x_c(t) = x_c(0)+\int_0^t u_c(s)ds,
\end{array}
%\right.
\end{equation}
and 
\begin{equation}\label{eq:piston-r}
%\left\{
\begin{array}{ll}
\dfr{\pt \bW}{\pt t} +\dfr{\pt\bF(\bW) }{\pt x} =0,  &  -\iy<x< x_c(t), \ \ t>0,\\[3mm]
\bW(x,0)=\bW_-(x),& -\iy<x< x_c(0). \\[3mm]
u(x_c(t),t)=u_c(t), & x_c(t) = x_c(0)+\int_0^t u_c(s)ds,
\end{array}
%\right.
\end{equation}
where the velocity $u_c(t)$ is defined through \eqref{eq:dynam-piston}. 
%The boundary conditions in \eqref{eq:piston-l} and \eqref{eq:piston-r} are the one-dimensional version of \eqref{eq:bc-2d}.
%[23:30, 28 Apr.]Obviously, the motion of the rigid body or the piston is determined by its position and the velocity, which are further determined by the force exerted from the opposite side. 
Note that the pressure gradient determines the acceleration and  the pressure exerted on the piston is not identical in general. 
Therefore, $u_c(t)$ is not constant in $t$ and the piston is accelerated. This is called {\em the accelerated piston problem.}  This observation will be put in the design of the high order  MBT method in this paper.

\subsection{Moving boundary tracking methods}\label{sec:mbt-framework}

The high order moving boundary  tracking (MBT) method we are going to propose   works in the finite volume framework, along with the cell-merging algorithm.  Since the control volumes are Eulerian in the interior of the computational domain, we just focus on  boundary control volumes. For easy understanding of the presentation,  we first describe the  one-dimensional version and then the two-dimensional case. 
\vspace{0.2cm}

In one dimension, we consider the piston problem from the left and assume that the fluid occupied region is $\bigcup_{0<t<T}[-\infty, x_c(t)]$ by ignoring the influence of the far field in the left.  The computational domain is divided into uniform cells
\begin{equation}\label{eq:mesh-1d}
\mathcal{T}=\bigcup_j\{I_j=(x_{j-\frac 12},x_{j+\frac 12}) : x_{j+\frac 12}=(j+\frac 12)\Dx, \ j\in\mathbb{Z}\}. 
\end{equation} 
We focus on the moving boundary  (piston) cell. Assume that the piston is located in the rightmost cell $I_J$ indexed  by $J$, i.e.  $x_c(t_n)\in I_J$. Denote by $I^c_{J}(t)=(x_{J-\frac 12},x_c(t))$ the ``legal"  cell according to the following cell-merging criterion. Then we apply \eqref{eq:euler} over the space-time control volume $B_J(t_n,t_{n+1}):=\{(x,t); x\in I_J^c(t), t_n\leq t\leq t_{n+1}\}$, 

\begin{equation}\label{eq:balance-1d}
\bga{rr}
%\d  \int_{I^\Ftext(t_{n+1})}\bW(x,t_{n+1})dx=\int_{I^\Ftext(t_n)}\bW(x,t_n)dx\\[3mm]
%\ \ \ \ \ \ \ \ \ \ \ \ \ \ \ \d   - \left\{\int_{t_n}^{t_{n+1}}\Big[\bF(\bW(x_\text{p}(t),t))-\dot{x}_\text{p}(t)\bW(x_\text{p}(t),t)\Big]dt
% - \int_{t_n}^{t_{n+1}}\bF(\bW(x_{j-\frac 12},t))dt\right\},
\d  \int_{I_J^c(t_{n+1})}\bW(x,t_{n+1})dx&=\d\int_{I_J^c(t_{n})}\bW(x,t_{n})dx - \left\{\int_{t_n}^{t_{n+1}}\Big[\bF(\bW(x_c(t)-0,t))-u_c(t)\bW(x_c(t)-0,t)\Big]dt\right.\ \ \ \\[3mm]
&  \left.\d- \int_{t_n}^{t_{n+1}}\bF(\bW(x_{J-\frac 12},t))dt\right\},
\eda
\end{equation}%\frac{1}{\sqrt{1+\dot{x}_\text{p}(t)}}
where $\bW=(\rho,\rho u,\rho E)^\top$ and $\bF(\bW)=(\rho u, \rho u^2+p, u(\rho E+p))^\top$.  The MBT method consists of three ingredients: {\em the flow evolution, the boundary tracking and the cell merging}.  As usually implemented for the flow evolution, we need to approximate the fluxes properly. The interior  flux $ \int_{t_n}^{t_{n+1}}\bF(\bW(x_{J-\frac 12},t))dt$ is evaluated using the standard GRP solver \cite{Li-1}. However,  the boundary flux  depends on the  one-sided GRP solver that provides the values
\begin{equation}
\bW_{c,-}^{n,*} :=\lim_{t\rw t_n+0} \bW(x_c(t)-0,t), \ \ \ \  \Big(\dfr{d\bW}{dt}\Big)_{c,-}^{n,*} :=\lim_{t\rw t_n+0} \left(\frac{d_c\bW}{dt}\right)(x_c(t)-0,t),
\label{value:GRP}
\end{equation}  
where $d_c/dt=\pt/\pt t+u_c(t)\pt/\pt x$ is the directional derivative along the boundary $x=x_c(t)$.  This pair of values also serves to track the boundary, as expressed in \eqref{eq:b-t}. 
\vspace{0.2cm}

As the piston travels to the next time level $t=t_{n+1}$,  the boundary cell $(x_J, x_c(t_{n+1}))$ could be very small or large so that  it should be redistributed, obeying a  cell-merging criterion described below. 
\vspace{0.2cm}

{\n\bf 1-D Cell Merging Criterion (CMC)}

\begin{enumerate}
\item[a.]  If $x_c(t_{n+1})\in I_J$ and  $|I_J^c(t_{n+1})|>\kappa\Dx$  where $\kappa$ is a user-tuned  parameter,  then $I_J^c(t_{n+1})$ is still well-defined.   Otherwise, $I_{J-1}$ and $I_J^c(t_{n+1})$ are merged to form a new boundary cell, denoted as $I_{J-1}^c(t_{n+1})$.

\item[b.] If $x_c(t_{n+1}) \in I_{J+1}$, the boundary control volume is  $I_J^c(t)=(x_{J-\frac 12}, x_c(t_{n+1}))$.

\item[c.] If $x_c(t_{n+1}) \in I_{J-1}$, the boundary cell is $I_{J-1}^c(t_{n+1})=(x_{J-\frac 32}, x_c(t_{n+1}))$.\\
\end{enumerate}

\vspace{2mm}
In two dimensions,  we still use the Cartesian meshes, 
\begin{equation}\label{eq:mesh-2d}
\mathcal{T}=\bigcup_{jk}\{I_{jk}=(x_{j-\frac 12},x_{j+\frac 12})\times(y_{k-\frac 12},y_{k+\frac 12}) : x_{j+\frac 12}=(j+\frac 12)\Dx, \ y_{k+\frac 12}=(k+\frac 12)\Dy, \ j, k\in\mathbb{Z}\}.
\end{equation}
\begin{figure}[!htb]
\centering
\includegraphics[width=.8\linewidth]{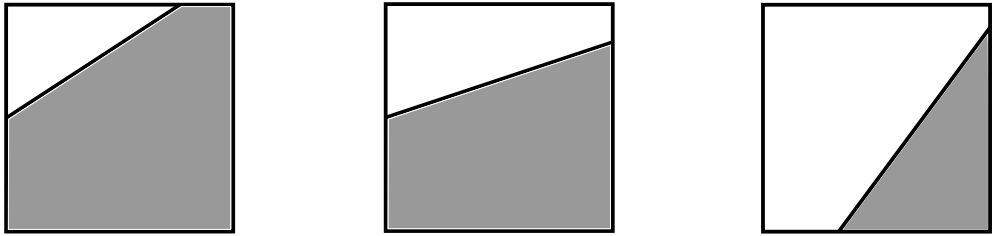}
\caption[small]{A cut cell may be either a triangle, a quadrangle or a pentagon: The shaded parts represent the solid body $\Om(t)$.}
\label{fig:cut-cell}
\end{figure} 
The Cartesian computational cells are usually cut by the boundary $\pt\Om$ of  solid objects to form very small ``illegal" cells.  We use the cell merging approach to modify boundary cells \cite{quirk}. Generally, the shape of a cut cell may be a triangle, a quadrangle or a pentagon, as shown in Figure \ref{fig:cut-cell}.  For the time being, we assume $I_{JK}^c$ is a ``legal" boundary cell surrounded by  the interior interface $\Gm_{int}(t)$  and the moving boundary $\Gm_b(t)$. Denote by $B_{JK}(t_n,t_{n+1}) =\{(x,y,t); (x,y)\in  I_{JK}^c(t), t_n\leq t\leq t_{n+1}\}$  the  space-time control volume. As the flow equations \eqref{eq:euler} are applied over this control volume, we obtain 
\begin{equation}
\bga{l}
\d\int_{I_{JK}^c(t_{n+1})} \bW(\bx,t_{n+1})d\bx = \int_{I_{JK}^c(t_{n})}\bW(\bx,t_n)d\bx\\[3mm]
\d \ \ \ \ \ \ \ \ \ \ \ \ \ \ \ \ \ \ \ \ \ \ \ \ \ \ \ \ \ \ \  \ \ \ -\int_{t_n}^{t_{n+1}} 
\left\{ \int_{\Gm_{int}(t)} (\bF,\bG)^\top\cdot\bn dldt + \int_{\Gm_{b}(t)} \big[(\bF,\bG)^\top\cdot\bn-(\bu\cdot\bn)\bW\big] dldt \right\},
\eda
\label{eq:2fv} 
\end{equation}
where $\bn$ is the unit outer normal vector of $\pt I_{JK}^c(t)$.
In parallel to the one-dimensional case, we develop 2-D one-sided GRP solver to provide the instantaneous values 
\begin{equation}
\bW_b^{n,*} =\lim_{t\rw t_n+0} \bW(\bx_b(t),t),\ \ \ \  \Big(\frac{d\bW}{dt}\Big)_b^{n,*}=\lim_{t\rw t_n+0} \dfr{d_c \bW}{dt}(\bx_b(t),t),
\label{value:grp-1d}
\end{equation} 
where $\bx_b(t_n)$ is any point on $\Gm_b$ at $t=t_n$, and $d_c/dt = \pt/\pt t+ u_c\pt/\pt x+ v_c\pt/\pt y$.  These values  serve to approximate the flux and the track the moving boundary  in the spirit of the GRP method.
\vspace{0.2cm} 

The tracking  of the moving boundary $\Gm_b$ is much more involved, compared to the one-dimensional counterpart, and 
is described by \eqref{eq:motion}.
%will be postponed   in Section \ref{sec:mbt-fv} for the algorithm description.  
For the efficiency of algorithm,  the cell-merging procedure is still necessary to avoid very small boundary cells. 
\vspace{0.2cm}

The cell-merging procedure is stated briefly as follows. 
Suppose that  the volume of $I_{JK}^c(t)$ is small enough, i.e.,  $|I_{JK}^c(t)|<\kappa\Dx\Dy$,  $\kappa$ is a user-tuned parameter. Then  it should be absorbed by its neighbors to form a larger ``legal" computational cell.  Denote the normal vector at the segment $\pt\Om(t)\cup I_{JK}$ by $\bn_{JK}(t)=(n^x_{JK}(t),n^y_{JK}(t))^\top$.
\vspace{0.2cm}

{\n\bf 2-D Cell Merging Criterion (2D-CMC)}

\begin{figure}[!htb]
\centering
\subfigure[The cell merging of a triangle]{
\includegraphics[width=.28\linewidth]{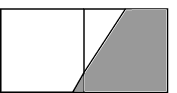}
\put(-40, 30){$I_{JK}$}
\put(-110, 30){$I_{J-1,K}$}
\label{subfig:cell-merge-a}
}
\ \ \ 
\subfigure[The cell merging  of a quadrangle]{
\includegraphics[width=.28\linewidth]{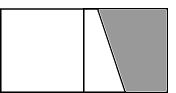}
\put(-40, 30){$I_{JK}$}
\put(-110, 30){$I_{J-1,K}$}
\label{subfig:cell-merge-b}
}
\ \ \
\subfigure[The cell merging  of a pentagon]{
\includegraphics[width=.28\linewidth]{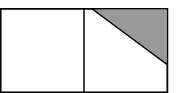}
\put(-40, 30){$I_{JK}$}
\put(-110, 30){$I_{J-1,K}$}
\label{subfig:cell-merge-c}
}
\caption[small]{The cell merging procedure.}
\label{fig:cell_merge}
\end{figure}

\begin{enumerate}
\item[a.] If $I_{JK}^c(t_n)$ is a triangle, combine it with its neighbor. For example, consider the situation shown in Figure \ref{subfig:cell-merge-a}, where $|n^x_{JK}(t_n)|>|n^y_{JK}(t_n)|$, combine $I_{JK}$ with $I_{J-1,K}$. Otherwise,  combine $I_{JK}$ with $I_{J,K+1}$.

\item[b.] If $I_{JK}^c(t_n)$ is a quadrangle, combine it with its uncut neighbor. For example, consider the situation shown in Figure \ref{subfig:cell-merge-b}, combine $I_{JK}$ with $I_{J-1,K}$.

\item[c.] Consider the situation that the  cut cells are pentagons. 
Assume that a pentagon has two cut edges. Then  there are two cases for the merging,   depending on their lengths.
\begin{enumerate}
\item[(i) ] If one of its cut edges is  too short, 
combine the pentagon with its neighbor to lengthen this cut edge, as shown in Figure \ref{subfig:cell-merge-c}.

\item[(ii)] Otherwise, leave the pentagon as a ``legal" cell.
\end{enumerate}
  
\end{enumerate}

%The cell-merging procedure is performed to combine $I_{JK}$ and $I_{J+1,K}$.
%Other delicate cases will be left in Subsection \ref{sec:mbt-2d-fv} for the implementation of the algorithm. 

%The key element of the moving boundary tracking method is to calculating $\bP_{c}^{n+\frac 12}$ in \eqref{eq:balance-bcv-general} and $PQ_{\Gm^{(0)}}^{n+\frac 12}$ in \eqref{eq:balance-2d}, which is the topic of the next section.\\

\vspace{2mm}
\section{One-sided generalized Riemann problem (OS-GRP) solver}\label{sec:solver}
In order to develop a  high order 
moving boundary tracking scheme in the finite volume framework, we need  to approximate the flux with a high order accuracy, which boils down to solving the following   one-sided generalized Riemann problem (GRP). In this sense, the accelerated piston problem \eqref{eq:piston-l} is also called {\em the one-sided generalized Riemann problem}. 
Two cases are discussed: the 1-D one-sided GRP solver and the 2-D one-sided GRP solver. The difference between them is that the transversal effect relative to the boundary interface is incorporated in the 2-D GRP solver.

\subsection{One-sided generalized Riemann problem solver in one dimension (OS-GRP-1D)}\label{sec:mbt-1d-1}
The one-sided GRP for the Euler equations in one space dimension is formulated as
\begin{equation}\label{eq:os-grp-1d}
%\left\{
\begin{array}{ll}
\dfr{\pt \bW}{\pt t} +\dfr{\pt\bF(\bW)}{\pt x}  =0, \ \ \ & -\iy<x<x_c(t), \ t>0,\\[3mm]
\bW(x,0) =\bW_-(x), & -\iy<x<0,\\[3mm]
u(x_c(t),t)=u_c(t), & x_c(t) = x_c(0)+\int_0^t u_c(s)ds.
\end{array} 
%\right.
\end{equation} 
where $ \bW_-(x)$ is a smooth vector function and usually taken as a polynomial for the numerical purpose.  The boundary trajectory $x=x_c(t)$ is defined following the dynamical system as
\begin{equation}\label{eq:boundary-motion}
\begin{array}{ll}
\dfr{dx_c(t)}{dt} =u_c(t),\ \ \ \ \\[3mm]%:= u(x_c(t) t))
\dfr{du_c(t)}{dt} =- \mathcal{S}_c\dfr{p(x_c(t)+0,t)-p(x_c(t)-0,t)}{\mathcal{M}_c}, 
\end{array}
\end{equation} %the sectional area is taken as $\mathcal{S}_c=1$, 
where the pair of the initial position and  velocity of the piston $(x_c(0), u_c(0))=(0, u_0)$ is prescribed, and the pressures $p(x_c(t)\pm 0),t)$ will be given later. Compared to \eqref{eq:dynam-piston}, the thickness $L$ of the piston is assumed to be infinitely small and ignored. 

\begin{rem}
This one-sided generalized Riemann problem can be formulated from the right-hand side $x_c(t)<x<\iy$ in the same way, thanks to the Galilean invariance.  
\end{rem} 
 
Such a problem  can be solved rigorously at least for a short time, following \cite{Takeno-1995}. However, we are satisfied with the calculation of the instantaneous values 
\begin{equation}
\bW_{c,-}^* =\lim_{t\rw 0+} \bW(x_c(t)-0,t), \ \ \Big(\dfr{d \bW}{dt}\Big)_{c,-}^*=\lim_{t\rw 0+} \dfr{d_c}{dt} \bW(x_c(t)-0,t), 
\tag{\ref{value:GRP}}
%\label{OS-GRP-1} 
\end{equation}
for the implementation of the high order  moving boundary tracking method we propose.   The same as the general GRP methodology \cite{Li-1},  the calculation of  \eqref{value:GRP} depends on solving the associated one-sided Riemann problem as formulated below. \\

\n{\bf Associated one-sided Riemann solver.} In order to solve the above one-sided GRP \eqref{eq:os-grp-1d},  
%consider the one-dimensional Euler equations in \eqref{eq:euler}. 
we firstly solve the one-sided Riemann problem following \cite{Courant-Friedrichs}. The associated one-sided Riemann problem is  defined by approximating \eqref{eq:os-grp-1d} with a constant initial data and neglecting the acceleration of the piston, which leads to the following IBVP
\begin{equation}
\label{data:OSRP-l}
%\left\{
\begin{array}{ll}
\dfr{\pt\bW^A}{\pt t} + \dfr{\pt\bF(\bW^A)}{\pt x}  = 0, &-\iy<x<x_c^A(t), t>0,  \\[2.5mm]
\bW^A(x,0) =\bW_-,  &  -\iy<x<0, \\[3mm]
u^A(x^A_c(t),t) \ev u_c(0), & x_c^A(t)=u_c(0)  t,
\end{array}
%\right.
\end{equation} 
where $\bW_-$ is regarded as a constant state.
%[11:00, 29 Apr.]The one-sided Riemann problem is solved in the region $\{(x,t): -\iy<x<x_c(t), t>0\}$.  
The boundary condition in \eqref{data:OSRP-l} says that the piston moves with a uniform velocity, which alludes to the fact that no force is exerted on the piston.  This is very different from the GRP solver that exhibits the acceleration effect. 
\vspace{0.2cm}
 
\begin{figure}[!htb]\label{Fig:grp-rp-1d}
\centering
\subfigure[The Riemann problem]{
\includegraphics[width=.35\linewidth]{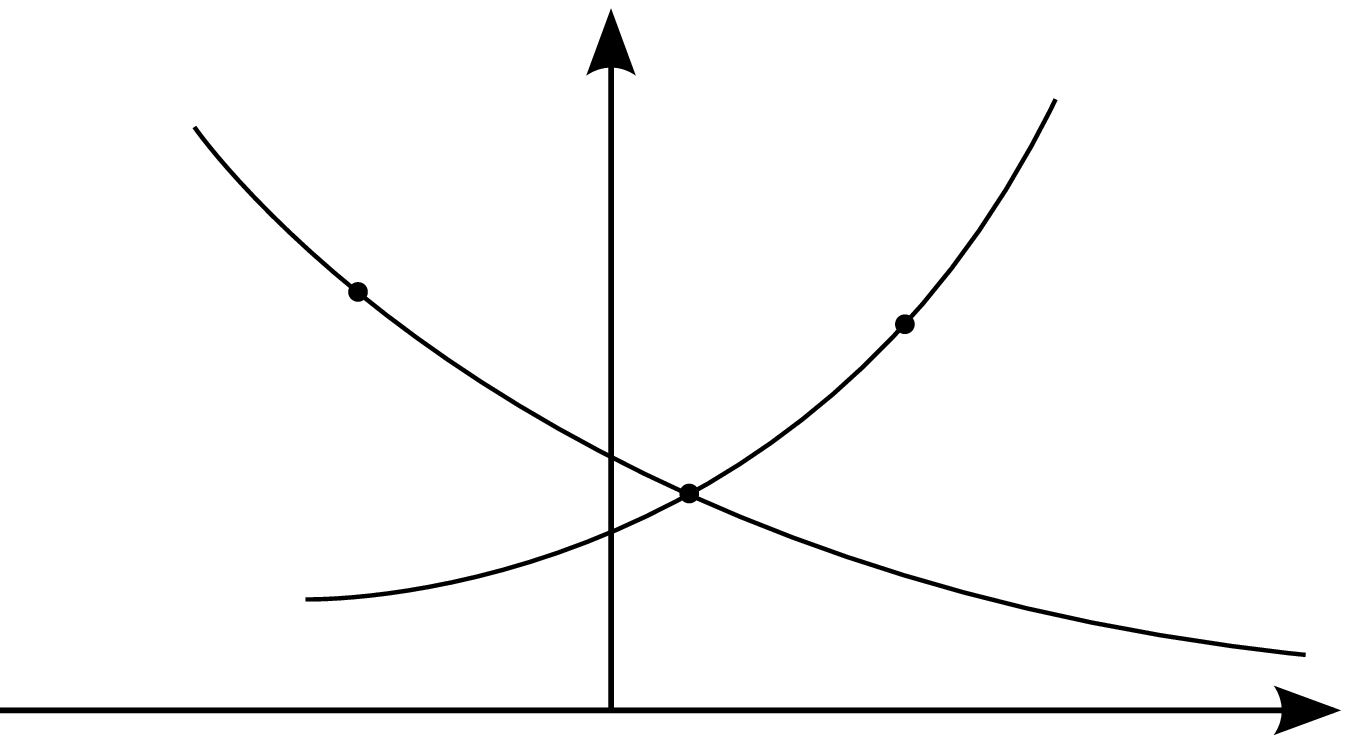}
\put(-85, 98){$ p $}
\put(0, 8){$ u $}
\put(-125, 68){$ \bW_- $}
\put(-55, 50){$ \bW_+ $}
\put(-86, 26){$ \bW^{*} $}
\label{subfig:riemann-0}
}
\ \ \ \ \
\subfigure[The classical piston problem]{
\includegraphics[width=.35\linewidth]{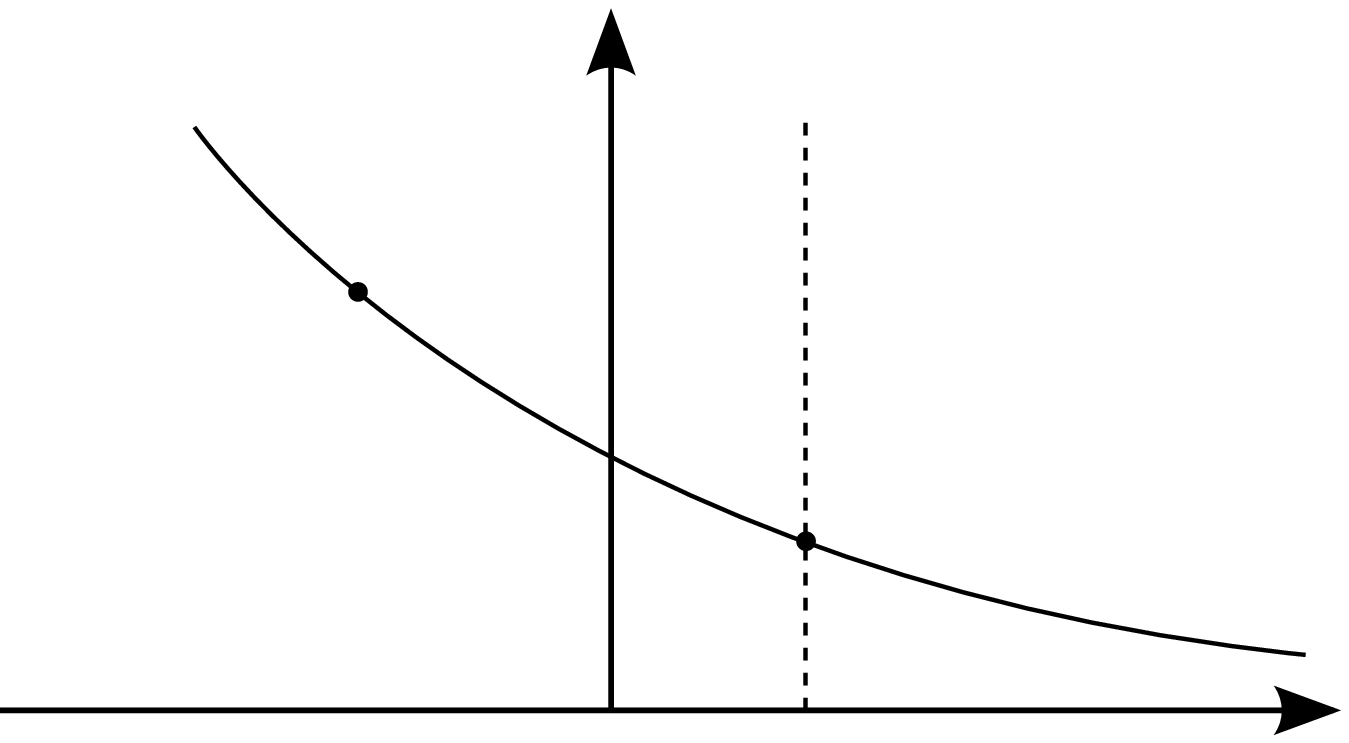}
\put(-85, 98){$ p $}
\put(0, 8){$ u $}
\put(-125, 68){$ \bW_- $}
\put(-63, 38){$(\bW^A) _{c,-}^*$}
\put(-70, 3){$u_c(0)$}
\label{subfig:riemann-half-0}
}
\caption[small]{The state curves of the standard Riemann problem and the one-sided Riemann problem in the $(u,p)$-plane.}
\label{fig:riemann-0}
\end{figure}

The one-sided Riemann solver follows from the standard Riemann solver, e.g. in \cite{mbt-falc}, to obtain the one-sided Riemann solution $(\bW^A)_{c,-}^*$, by noting that the velocity of the one-sided Riemann solution is
\begin{equation}
(u^A)_{c,-}^*=u_c(0).
\end{equation} 
The pressure $(p^A)_{c,-}^*$ exerted on the piston is obtained through the standard analysis for the Riemann solver, as figuratively shown in Figure \ref{Fig:grp-rp-1d}. %Details are given in Appendix \ref{app:piston-polytropic}. 

%, which corresponds to the acceleration calculation \eqref{eq:acc-1d}.\\
\vspace{0.2cm}
%\begin{rem}
%The above procedure is the standard method to resolve waves in the Riemann problem of the one-dimensional Euler equations.
%Therefore the classical piston problem is also called the one-sided Riemann problem in \cite{mbt-falc}. The conventional Riemann problem is solved by calculating the intersection point of the two state curves passing the two initial state as shown in Figure \ref{subfig:riemann-0}. In the one-sided Riemann problem,  only the left initial state $\bW_-$ is available. However, the velocity of the fluid is given by $u_c$ due to the boundary condition in \eqref{eq:ibvp-RP}. This is figuratively shown in Figure \ref{subfig:riemann-half-0}.
%\end{rem}
%the $u=u_c$ on one side for a given state. The details will be given in Appendix ???. Appendix \ref{app:piston-polytropic}.\\

\n{\bf One-sided GRP solver (OS-GRP-1D).}  The one-sided GRP \eqref{eq:os-grp-1d} is asymptotically consistent with the one-sided Riemann problem \eqref{data:OSRP-l} in the sense that 
\begin{equation}
\bW^*_{c,-}= (\bW^A)^*_{c,-}. 
\end{equation} 
Symmetrically, the one-sided GRP from the right-hand side, as formulated in \eqref{eq:piston-r},  is associated with the right one-sided Rimann problem,  
\begin{equation}
\bW^*_{c,+}=(\bW^A)_{c,+}^*,
\end{equation} 
where the annotation of obvious notations is suppressed. Note that the pressures from both sides are different, generally speaking 
\begin{equation}
p^*_{c,-} \neq p^*_{c,+}.
\end{equation}
So the piston is accelerated, according to \eqref{eq:boundary-motion}, 
\begin{equation}
\left.\dfr{du_c(t)}{dt}\right|_{t=0} = -\mathcal{S}_c \dfr{p^*_{c,+}-p^*_{c,-}}{\mathcal{M}_c}.
\end{equation}
This is the instantaneous acceleration of the piston for the one-sided GRP solver.  Once we have this, we immediately use the standard GRP solver \cite{Li-1} to derive the relation 
\begin{equation}\label{eq:linear-GRP}
\bga{l}
a^* \Big(\dfr{d u}{d t}\Big)_{c,-}^* + b^*\Big(\dfr{d p}{d t}\Big)_{c,-}^* = d^*,  
\eda
\end{equation}
where $\frac{d}{dt}=\frac{\pt}{\pt t}+u\frac{\pt}{\pt x}$,  the coefficients $ a^* $, $ b^* $ and $ d^* $ are determined from  $\bW_-(0)$, $\bW_-'(0)$ and $\bW_{c,-}^*$, as usually done for the GRP solver \cite{Li-1}. 
The procedure to calculate coefficients in \eqref{eq:linear-GRP} exactly follows that given in \cite{Li-1}. Readers are referred there for details.

According to the boundary condition in \eqref{eq:os-grp-1d}, we have $(\frac{d u}{d t})_{c,-}^*=\frac{du_c}{dt}(0)$. Then $(\frac{d p}{d t})_{c,-}^*$ is obtained by solving the linear equation \eqref{eq:linear-GRP}.
As for the density, we have from the EOS, thanks to the fact  $ (ds/d t)^*_{c,-}=0$, 
\begin{equation}\label{eq:GRP-rho}
\Big(\dfr{d\rho}{d t}\Big)_{c,-}^* = \dfr{1}{{(c^*)}^2}\Big(\dfr{d p}{d t}\Big)_{c,-}^*,
\end{equation}
where $c^*$ is the local sound speed defined by $c^*=\sqrt{\gm p^*_{c,-}/\rho^*_{c,-}}$.

%thanks to the facts that $ \Big(\dfr{d_c s}{d t}\Big)_*=0$ and $\frac{\pt p}{\pt \rho}(\rho,s)=c^2$. 

\vspace{2mm}

\subsection{One-sided generalized Riemann problem solver in  two dimensions (OS-GRP-2D)}\label{sec:mbt-1d-2}

Think of  a solid body  in two dimensions with  the initial translational and rotational velocities $\bu_c(0)$ and $\om(0)$.  The pressure exerted on the surface determines the acceleration. Our strategy is made as follows: At any point on the surface, we will first solve the one-sided Riemann problem normal to the surface to obtain the pressure, which yields the translational and rotational accelerations %total pressure 
of the surface according to \eqref{eq:2d-translational} and \eqref{eq:2d-rotational}.  Then we develop 2-D one-sided GRP solver to obtain the instantaneous value of  the derivative  $d\bW/dt$ for the flux evaluation and the boundary tracking. \vspace{0.2cm}

Denote  by $\pt\Om(0)$ the initial position of the solid surface and recall the notations in \eqref{eq:motion}. For any $\bX_b\in\pt\Om(0)$, denote by $\bx_b(t;\bX_b)$ the  trajectory from $\bX_b$ for $t>0$.
%, the detailed calculation of which will be given later in Subsection \ref{sec:mbt-2d-fv}. 
The unit normal vector of $\pt\Om(t)$ at $\bx_b(t;\bX_b)$ is denoted by $\bn(t;\bX_b)$ and the instantaneous velocity of the solid surface at $\bx_b(t;\bX_b)$ is denoted by $\bu_b(t;\bX_b)$.

 \vspace{0.2cm} 
 
\n{\bf  One-sided normal Riemann solver.}  
For the sake of presentation simplicity, we place the section of the moving boundary we concern initially along the $y$-axis, thanks to the Galilean invariance.
Define the associated one-sided normal Riemann problem at any point $(0,y_*)$ on the surface of the solid body as
\begin{equation}
\label{data:OSRP-II}
%\left\{
\begin{array}{ll}
\dfr{\pt\bW^N}{\pt t} + \dfr{\pt\bF(\bW^N)}{\pt x} = 0, &-\iy<x<x_c^N(t), \ t>0, \\[2.5mm]
\bW^N(x,t=0) =\bW_-^N(0,y_*),  &  -\iy<x<0, \\[3mm]
%{\color{red}\bu^N(x_b(t),t) \ev \bu_b(0),} & {\color{red}\bx_b(t)=(0,y_*)+\bu_b(0) \ t \ \text{(This is wrong.)}}, \ t>0
u^N(x^N_c(t),t) \ev u_c^N, & x^N_c(t)=u_c^N \ t, \  t>0.
\end{array}
%\right.
\end{equation} 
where the superscript ``$N$" indicates the vector normal to the boundary $x=0$. 
%where $\bu_b(0)=\bu_b(0,y_*,0)$ and the $x$-axis can be regarded as the normal of the surface. 
As far as the associated one-sided normal Riemann problem is concerned, the piston velocity in the IBVP \eqref{data:OSRP-II} is defined as $u^N_c=\bu_b(t=0;0,y_*)\cdot\bn(t=0;0,y_*)$, where $\bu_b(t=0;0,y_*)$ is given by the first equation in \eqref{eq:motion}.
The IBVP \eqref{data:OSRP-II} can be solved exactly in the same way as the one-dimensional one-sided Riemann problem \eqref{data:OSRP-l}.  
% This o
The corresponding result of solving the above associated one-sided normal Riemann problem is denoted as $(\bW^N)^*$.
\vspace{0.2cm}

%{\red We should put the genuinely 2-D piston problem \eqref{eq:os-grp-2d} at the beginning of this subsection. Then associated one-sided RP. Then homogeneous one-sided GRP.}\\

\n{\bf One-sided GRP solver in 2-D.}  Once we know the total pressure on the surface  using the one-sided normal Riemann solver, we can calculate the acceleration according to \eqref{eq:2d-translational} and \eqref{eq:2d-rotational}, the numerical discretizations of which will be specified in Subsection \ref{sec:mbt-2d-fv}. Then we can solve the following two-dimensional one-sided generalized Riemann problem,  
\begin{equation}\label{eq:os-grp-2d}
%\left\{
    \begin{array}{ll}
\dfr{\pt\bW}{\pt t} + \dfr{\pt\bF(\bW)}{\pt x} + \dfr{\pt\bG(\bW)}{\pt y} = 0, \\[2.5mm]% x<x_\ptext(y,t), \\[2.5mm]
\bW(x,y,0)=\bW_-(x,y), \ &x<0, \  y\in\mathbb{R},\\[2.5mm]
%{\red \bu(\bx_b(t),t)\ =\bu_b(\bx_b(t),t),}  &  t>0,
\bu(\bx_b(t;0,y),t)\cdot\bn(t;0,y)\ =\bu_b(t;0,y)\cdot\bn(t;0,y),  & y\in\mathbb{R}, \  t>0,
    \end{array}
%\right.
\end{equation}
where, in light of \eqref{eq:motion}, $\bx_b(t;0,y)$ is the trajectory of the boundary point starting from $(0,y)$ and its velocity is $\bu_b(t;0,y)$.
%Let $\bx_b(t;\bX_b)$ be the trajectory of the motion starting  from $\bX_b=(0,y_*)\in\pt\Om(0)$.  
The 2-D GRP solver serves to obtain the instantaneous values  at any point $(0,y_*)$
\begin{equation}
\bW_b^*=\lim_{t\rw 0+}\bW(\bx_b(t;0,y_*),t), \ \ \ \ \Big(\dfr{d\bW}{dt}\Big)_b^*=\lim_{t\rw 0+}\dfr{d_c\bW}{dt} (\bx_b(t;0,y_*),t),
\end{equation}
for any $y_*\in \Re$ and particularly $y_*=0$, where $\frac{d_c}{dt}=\frac{\pt}{\pt t}+u\frac{\pt}{\pt x}+v\frac{\pt}{\pt y}$ is the directional derivative along the particle path.
Remind again that $\bW_b^*$ is just the one-sided normal Riemann solution of \eqref{data:OSRP-II}, and is given as $\bW_b^*=(\bW^N)^*$, which provides the local acceleration.

\vspace{0.2cm}

Different from the one-dimensional one-sided GRP solver, the transversal effect,  as pointed out in \cite{Lei-Li-2019-AMM}, is important. Following \cite{Du-Li-1},  we regard the transversal term as a local source term and solve the following IBVP,
\iffalse
\begin{equation}\label{eq:os-grp-2d-quasi}
%\left\{
    \begin{array}{ll}
\dfr{\pt\bW}{\pt t} + \dfr{\pt\bF(\bW)}{\pt x} = -\Big(\dfr{\pt\bG}{\pt y}\Big)^*, & x<x_c(t) ,\\[2.5mm]% x<x_\ptext(y,t), \\[2.5mm]
\bW(x,t=0)=\bW_-(x,y_*), & -\infty<x<0,\\[2.5mm]
%{\red \bu(x_b(t),t) = \bu_b(t; 0,y_*)}, & \ t>0,
u(x_c(t),t)=u_c(t),  & x_c(t)=\int_0^tu_c(s)ds \ t>0,
    \end{array}
%\right.
\end{equation}
where $(\frac{\pt\bG}{\pt y})^* =\frac{\pt \bG}{\pt \bW}(\bW_b^*) (\frac{\pt\bW}{\pt y})^* $ is a fixed value, as described in \cite{Du-Li-1}.  
The piston velocity $u_c(t)$ involved in \eqref{eq:os-grp-2d-quasi} is given as $u_c(t)=u_c(0)+a_c(0)t$, where
\begin{equation}
u_c(0)=\bu_b(t=0;0,y_*)\cdot\bn(t=0;0,y_*), \ \ \ a_c(0)=\dfr{d\bu_b}{dt}(t=0;0,y_*)\cdot\bn(t=0;0,y_*).
\end{equation}
\fi
\begin{equation}\label{eq:os-grp-2d-quasi}
    \begin{array}{ll}
\dfr{\pt\bW}{\pt t} + \dfr{\pt\bF(\bW)}{\pt x} = -\Big(\dfr{\pt\bG}{\pt y}\Big)^*,  \\[2.5mm]
\bW(x,y,0)=\bW_-(x,y), \ &x<0, \  y\in\mathbb{R},\\[2.5mm]
\bu(\bx_b(t;0,y),t)\cdot\bn(t;0,y)\ =\bu_b(t;0,y)\cdot\bn(t;0,y),  & y\in\mathbb{R}, \  t>0,
    \end{array}
\end{equation}
where $(\frac{\pt\bG}{\pt y})^* =\frac{\pt \bG}{\pt \bW}(\bW_b^*) (\frac{\pt\bW}{\pt y})^* $ is a fixed value, as described in \cite{Du-Li-1}.
Then the 2-D one-sided GRP solver for \eqref{eq:os-grp-2d} is derived exactly the same as the 2-D GRP solver in \cite{Li-2, Du-Li-1}. %As a final result, we have $\bW^*=(\bW^N)^*$ and $(\frac{d\bW}{dt})^*=(\frac{d\bW^N}{dt})^*$.

\vspace{2mm}
\section{High-order moving boundary tracking algorithm}\label{sec:mbt-fv}
This section describes a high order moving boundary tracking algorithm based the newly developed one-sided GRP solver.  This  resulting scheme is different from those using the ghost fluid approach \cite{Khoo-FSI-LevelSet, mbt-ilw}, but can be regarded as the high order extension  of MBT in  \cite{mbt-falc}.  

\subsection{One-dimensional high order MBT algorithm}\label{sec:mbt-1d-fv} The algorithm consists of three  ingredients, as pointed out in Section \ref{sec:preliminary}:  {\em The evolution of flows, the moving boundary tracking and the boundary cell merging. }  The flow motion in the interior domain is treated using  the standard finite volume method, e.g., the GRP method in \cite{Li-1}. So we focus on the boundary cell, which is denoted as $B_J(t_n, t_{n+1})=\{ (x,t); x\in  (x_{J-\frac 12}, x_c(t)), t_n\leq t\leq t_{n+1}\}$. The algorithm is described as follows. 
\vspace{0.2cm}

\begin{enumerate}

\item[(i)] {\bf Implementation of one-sided GRP solver.}  Given the initial data over the boundary cell $I_J^c$ at $t=t_n$, 
\begin{equation}
\bW_J^{c,n}(x) = \overline\bW^{c,n}_J+\Big(\frac{\pt\bW}{\pt x}\Big)^{c,n}_J\Big(x-\dfr{x_{J-\frac 12} + x_c(t_n)}{2}\Big),
\label{data:n}
\end{equation}
we use the one-sided GRP solver to obtain the instantaneous values 
$(\bW_{c,-}^{n,*},  (\frac{d\bW}{dt})_{c,-}^{n,*})$,  which is annotated in \eqref{value:GRP}. Then we compute the mid-point values for the flux approximation and interface values for slope evaluation at the next time level,
\begin{equation}\label{eq:boundary-value-1d}
\bW_{c,-}^{n+\frac 12}=\bW_{c,-}^{n,*}+\dfr{\De t}2 \Big(\dfr{d\bW}{dt}\Big)_{c,-}^{n,*},  \ \ \ \bW_{c,-}^{n+1}=\bW_{c,-}^{n,*}+\De t \Big(\dfr{d\bW}{dt}\Big)_{c,-}^{n,*}. 
\end{equation}
\vspace{0.2cm} 

\item[(ii)] {\bf The tracking of the moving boundary.}  We track the moving boundary to the next time level using the formula within the second order accuracy,
\begin{equation} 
{x_c}(t_{n+1})=x_c(t_n)+\Dt u_c^n+\dfr{\Dt^2}{2}\Big(\frac{d u_c}{dt}\Big)^n.
\end{equation} 

\item[(iii)] {\bf Flow evolution.} We use the finite volume framework \eqref{eq:balance-1d} to evolve the flow
\begin{equation}\label{eq:bcv-2}
\bga{l}
{|I_J^c(t_{n+1})|}\overline\bW^{c,n+1}_J={|I_J^c(t_n)|}\overline\bW^{c,n}_J- \Dt\left[\Big(\bF^{n+\frac 12}_{c,-}-u_c^{n+\frac 12}\bW^{n+\frac 12}_{c,-}\Big) - \bF^{n+\frac 12}_{J-\frac 12}\right],
\eda
\end{equation}
where $I_J^c(t_{n+1}) =(x_c(t_{n+1})-x_{J-\frac 12})$, $\overline{\bW}_J^{c,n+1}$ is the average of $\bW(x,t_{n+1})$ over $I_J^c(t_{n+1})$, and
\begin{equation}
\bF_{c,-}^{n+\frac 12} = \bF(\bW_{c,-}^{n+\frac 12}),  \ \ \ \ \bF_{J-\frac 12}^{n+\frac 12} = \bF(\bW_{J-\frac 12}^{n+\frac 12}).
\label{flux-1d} 
\end{equation} 
The half-time value $\bW_{J-\frac 12}^{n+\frac 12}$ is defined by $\bW_{J-\frac 12}^{n+\frac 12}=\bW^{n,*}_{J-\frac 12}+\frac{\Dt}{2}(\frac{\pt\bW}{\pt t})_{J-\frac 12}^{n,*}$, where $\bW^{n,*}_{J-\frac 12}$ and $(\frac{\pt\bW}{\pt t})_{J-\frac 12}^{n,*}$ are obtained by applying the standard GRP solver at $x_{J-\frac 12}$ \cite{Li-1}.
\vspace{0.2cm} 

\item[(iv)] {\bf Cell-merging procedure.}  Obey the 1-D cell-merging criterion to construct  a ``legal" boundary cell at the new time level $t=t_{n+1}$.
\vspace{0.2cm}

\item[(v)]  {\bf Data update.} Update the slope over  the boundary cell $I_J^c(t_{n+1})$,
\begin{equation}\label{eq:slope-bcv}
\Big(\dfr{\pt\bW}{\pt x}\Big)^{c,n+1}_J=\text{minmod}\left(\dfr{\bW^{n+1}_{c,-}-\bW^{n+1}_{J-\frac 12}}{|I_J^c(t_{n+1})|},\al\dfr{\overline\bW_J^{c,n+1}-\overline\bW_{J-1}^{n+1}}{(\Dx+|I_J^c(t_{n+1})|)/2}\right), 
\end{equation}
where $\al\in(0,2)$ is a user defined parameter. Then we reconstruct the data in the form \eqref{data:n} over the boundary cell $I_J^c(t_{n+1})$.

 \end{enumerate}
 \vspace{4mm}

 %%%%%%%%%%%%%%%%%%%%%%%%%%%%%%%%%%%
%%%%%%%%%%%%%%%%%%%%%%%%%%%%%%%%%%%
%[10:34, 7 May] If the linear reconstruction is adopted in the MBT scheme, one would expect the computation \eqref{eq:balance-bcv-general}  to be second-order accurate.

\iffalse
\begin{rem}
Similar analysis shows that, the MBT scheme developed in \cite{mbt-falc} is even zeroth-order accurate in boundary control volumes, since it only uses the one-sided Riemann solver and neglects the acceleration.
In a more general sense, the above analysis implies that any moving mesh method  may suffer the same problem in presence of the sudden jump of interface velocities.
\end{rem}

\begin{rem}
The decay of the numerical accuracy is caused by the sudden jump of the velocities of the two interfaces of $B_c(t)$, i.e. $x_c(t)$ moves with the velocity $u_c(t)$ while $x_{J-\frac 12}$ is fixed. 

Besides the one proposed above, another possible remedy for this defect is using the adaptive grid method and assure the condition $\Delta u=\mathcal{O}(\Dx)$ where $\Delta u$ denote the difference between velocities of the two neighboring interfaces.\\
\end{rem}
\fi

%The same argument tells us that the first-order scheme constructed in \cite{mbt-falc} is even non-consistent locally at the moving boundary.

%%%%%%%%%%%%%%%%%%%%%%%%%%%%%%%%%%%
%%%%%%%%%%%%%%%%%%%%%%%%%%%%%%%%%%%

 \vspace{2mm}
 \subsection{Two-dimensional high order MBT algorithm}\label{sec:mbt-2d-fv} The 2-D MBT algorithm has the same methodology as the  1-D case,  in addition that  the moving boundary tracking is more technical. Therefore we only provide details for boundary tracking and the data reconstruction, by assuming that all necessary values are available from the 2-D one-sided GRP solver.  We still use the Cartesian mesh as expressed in \eqref{eq:mesh-2d}. At $t_n$, denote by $I^c_\mu$  the cells cut by $\Om(t_n)$  for $\mu=1,2,\dots,M$. %, by $\bx_c(t_n)$ the barycenter. 
 Denote the segment of $\pt\Om(t_n)$  inside $I^c_\mu$ by $\Gm_\mu^n$ with the middle point $\bx_\mu^n$.
 \vspace{0.2cm} 
 
 \begin{enumerate}
\item[(i)] \textbf{2-D moving boundary tracking.}  In light of  \eqref{eq:motion}, the instantaneous velocity at    the middle point $\bx_\mu^{n}$  of $\Gm_\mu^{n}$  is given by
\begin{equation}
\bga{l}
\bu_b(\bx_\mu^{n},t_n)=\bu_c(t_n)+\om(t_n)\Big[\bx_\mu^{n}-\bx_c(t_n)\Big]^\perp.
\eda
\end{equation}
The associated one-sided normal Riemann problem \eqref{data:OSRP-II} is consequently defined at $\bx_\mu^n$.
Denote the corresponding one-sided Riemann solution by $\bW_{\bx_\mu^n}^{n,*}$. Particularly, the boundary pressure $p_{\bx_\mu^{n}}^{n,*}$ is obtained.
Therefore the translational and rotational accelerations  of $\Om(t_n)$ are, according to \eqref{eq:2d-translational} and \eqref{eq:2d-rotational}, 
\begin{equation}\label{eq:acceleration-2d}
\bga{l}
{\dfr{d\bu_c}{dt}}(t_n)=-\dfr{1}{\mathcal{M}_c} \sum\limits_{\mu}p_{\bx_\mu^{n}}^{n,*} \ |\Gm_\mu^{n}| \ \bn_\mu^{n}, \\[3mm]
{\dfr{d \om}{dt}}(t_n)=\dfr{1}{\mathcal{A}_c}\sum\limits_{\mu}\left\{p_{\bx_\mu^{n}}^{n,*} \ |\Gm_\mu^{n} | \  \left[ \bn_\mu^{n}\times  (\bx^{n}_\mu-\bx_c(t_n)) \right] \right\}.
\eda
\end{equation}
Consequently, in light of  \eqref{eq:motion}, the acceleration of the point $\bx_\mu^n\in\pt\Om(t_n)$ is
\begin{equation}
\bga{l}
\dfr{d\bu_b}{dt}(\bx_\mu^n,t_n)
=\dfr{d\bu_c}{dt}(t_n)+\dfr{d\om}{dt}(t_n)\big[\bx_\mu^{n}-\bx_c(t_n)\big]^\perp+\om(t_n)\Big\{\om(t_n)\big[\bx_\mu^{n}-\bx_c(t_n)\big]^\perp\Big\}^\perp.
\eda
\end{equation}
The above procedure of tracking the moving boundary provides the information of the piston motion required by the 2-D one-sided GRP solver to solve \eqref{eq:os-grp-2d-quasi} in Section \ref{sec:mbt-1d-2}.
At last, we update relevant values to the next time level as
\begin{equation}\label{eq:vel-evol}
\bga{l}
\bx_c(t_{n+1})=\bx_c(t_n)+\Dt\bu_c(t_n)+\dfr{\Dt^2}{2}\dfr{d\bu_c}{dt}(t_n),\\[2.5mm]
\theta(t_{n+1})=\theta(t_n)+\Dt\om(t_n)+\dfr{\Dt^2}{2}\dfr{d\om}{dt}(t_n),\\[2.5mm]
\bu_c(t_{n+1})=\bu_c(t_n)+\Dt\dfr{d\bu_c}{dt}(t_n),\ \ \om(t_{n+1})=\om(t_n)+\Dt\dfr{d\om}{dt}(t_n).
\eda
\end{equation}
\iffalse
Then the middle-point $\bx_\mu(t_n)$  is updated to
\begin{equation}\label{eq:position-estimation-2d}
\bx_\mu(t_{n+1})={\bx_c}(t_{n+1})+\bT(\Delta\theta)  \big[\bx_\mu(t_n)-\bx_c(t_n)\big],
\end{equation}
where 
\begin{equation}
\bT(\Delta \theta) = \left[
\bga{rr}
\cos(\Delta\theta) & -\sin(\Delta\theta)\\
\sin(\Delta\theta) & \cos(\Delta\theta)
\eda
\right],
\end{equation}
where $\Delta \th$ is
\begin{equation*}
 \Delta\theta=\theta(t_{n+1})-\theta(t_n)=\Dt\om(t_n)+\frac{\Dt^2}{2}{\dfr{d\om}{dt}}(t_n).
\end{equation*}
\fi

\item[(ii)] {\bf Data reconstruction over boundary cells. }  
For the cut cell $I^c_\mu(t_{n+1})$, denote all of its edges by $\Gm_{\mu}^{(k)}(t_{n+1})$ for $k=0,1,\dots,K$.
Furthermore, assume that $\Gm_{\mu}^{(0)}(t_{n+1})=\pt\Om(t_{n+1})\cap I_\mu(t_{n+1})$.
The unit outer normal vectors of $\Gm_{\mu}^{(k)}(t_{n+1})$ with respect to $I^c_\mu(t_{n+1})$ are denoted by $\bn_{\mu}^{(k)}(t_{n+1})$.
In order to linearly approximate $\bW(\bx,t_{n+1})$ in $I^c_\mu(t_{n+1})$, we need to estimate its gradient.
By the Gauss theorem, we have 
\begin{equation}\label{eq:gauss}
\int_{I^c_\mu(t_{n+1})}\nb\bW(\bx,t_{n+1})d\bx= \sum_{k=0}^K\int_{\Gm_{\mu}^{(k)}(t_{n+1})}\bW(\bx,t_{n+1})\otimes\bn(\bx,t_{n+1}) dl,
\end{equation}

To estimate the gradient of $\bW$ in $I^c_\mu(t_{n+1})$, we discretize  \eqref{eq:gauss} as
\begin{equation}
(\nabla\bW)_\mu^{n+1} = \dfr{1}{|I^c_\mu(t_{n+1})|}\sum_{k=0}^K{|\Gm_\mu^{(k)}(t_{n+1})|\bW_\mu^{(k),n+1} \otimes\bn_\mu^{(k),n+1}},
\end{equation}
where $\bW_\mu^{(0),n+1}=\bW^{n,*}_\mu+\Dt(\frac{d\bW}{dt})_\mu^{n,*}$, $\bW_\mu^{n,*}$ and $(\frac{d\bW}{dt})_\mu^{n,*}$ are obtained by solving the 2-D one-sided GRP at $(\bx_\mu^n,t^n)$.
%, following the procedure proposed in Subsection \ref{sec:mbt-1d-2}.
The least square limiter, which is a multi-dimensional version of \eqref{eq:slope-bcv}, can be used to suppress possible oscillations \cite{least_square_limiter}.
 
\vspace{2mm}

\end{enumerate}

\vspace{2mm}

\section{Accuracy analysis}\label{sec:accuracy_analysis}

In this section we check the accuracy of the proposed MBT method. We just focus on ``legal" boundary cells $I_J^c(t)$.   Recall the flux approximation in \eqref{flux-1d}. Then we estimate
\begin{equation}\label{eq:flux-error-inner}
 \d \bF_{J-\frac 12}^{n+\frac 12} - \dfr{1}{\Dt}\int_{t_n}^{t_{n+1}}\bF(\bW(x_{J-\frac 12},t))dt 
= -\dfr{\De t^2}{6}\dfr{\pt^2\bF}{\pt t^2} (x_{J-\frac 12},t_n) +\mathcal{O}(\De t^3),
\end{equation}
and 
\begin{equation}
\begin{array}{rl}
& [\bF_{c,-}^{n+\frac 12} -u_c^{n+\frac 12} \bW_{c,-}^{n+\frac 12}] -\dfr{1}{\De t} \int_{t_n}^{t_{n+1}}[\bF(\bW(x_c(t)-0,t))-u_c\bW(x_c(t)-0,t)]dt\\[3mm]
 =& -\dfr{\De t^2}6\dfr{d_c^2}{dt^2}[\bF(\bW)-u\bW](x_c(t_n)-0,t_n) +\mathcal{O}(\De t^3). 
\end{array}
\end{equation}
 The flux difference yields, 
\begin{equation}
\label{eq:truncation}
\bga{rl}
&\d\dfr{1}{\De t} \left\{\int_{t_n}^{t_{n+1}}\Big[\bF(\bW(x_c(t)-0,t))-u_c(t)\bW(x_c(t)-0,t)\Big]dt - \int_{t_n}^{t_{n+1}}\bF(\bW(x_{J-\frac 12},t))dt\right\}\\[3mm]
&\ \ \ \ \ \ \ \ \ \ \ \ \ \ \ \ \ \ \ \ \ \ \ \ \ \ \ \ \ \ \ \ \ \ \ \ \ \ \ \ \ \ \ \ \ \ \ \ \ \ \ \ \ \ \ \ \ \ \ \ \ \ \ \ \d- \left[\Big(\bF^{n+\frac 12}_{c,-}-u_c^{n+\frac 12} \bW_{c,-}^{n+\frac 12}\Big) - \bF^{n+\frac 12}_{J-\frac 12}\right]\\[3mm]
=&\dfr{\Dt^2}{6}\left[\dfr{d_c^2}{dt^2}(\bF(\bW)- u\bW)(x_c(t_n)-0,t_n)- \dfr{\pt^2}{\pt t^2} \bF(\bW)(x_{J-\frac 12},t_n)\right]+\mathcal{O}(\De t^3). 
\eda
\end{equation} 
It is evident that  the time varying of boundary cells introduces extra errors. If the boundary is fixed, i.e., $u_c\ev 0$,  we can achieve the second order accuracy for $\De t\sim \De x$, which is written as
\begin{equation}\label{eq:truncation-1}
\overline\bW^{n+1}_c-\dfr{1}{  |I_J^c(t_{n+1})|}\int_{I_J^c(t_{n+1})}\bW(x,t_{n+1})dx=\mathcal{O}(\Dt^3).
\end{equation}
Otherwise  the error has the form 
\begin{equation}
\label{eq:error-aver}
\ \ \ \overline\bW^{n+1}_c-\dfr{1}{|I_J^c(t_{n+1})|}\int_{I_J^c(t_{n+1})}\bW(x,t_{n+1})dx=\mathcal{O}(\Dt^3/\De x),
\end{equation}
which implies that if $\De t \sim \De x^{3/2}$, the second order accuracy can be also achieved. By the way,  if only the local one-sided Riemann solver is used \cite{mbt-falc}, the accuracy reduces to first order.  The numerical example below confirms such an observation. 
\vspace{0.2cm}

We use the example  proposed in  \cite{mbt-forrer} to verify the accuracy of the current MBT method, with the comparison with the original MBT just using the  Riemann solver  in \cite{mbt-falc}.   This example describes a tube filled with gas initially 
\begin{equation*}
\left[
\bga{l}
\rho_0(x)\\[3.5mm]
u_0(x)\\[3.5mm]
p_0(x)
\eda
\right]=\left[
\bga{l}
1+\dfr 15 \cos(2\pi x)\\[2.5mm]
2(x+0.5)\ u_c(0)\\[2.5mm]
{[\rho_0(x)]}^\gamma
\eda
\right], \ \ x\in(-0.5,0),
\end{equation*}
where $\gamma=1.4$ and $u_c(0)$ will be specified later.
The left and right ends of the tube is closed by a fixed wall and a moving piston, respectively.
Two  cases are considered.  The first case has 
 a uniform piston velocity $u_c(t)\equiv 0.5$, while the second case has a varying piston velocity $u_c(t)=2t$. %A remedy in practical computations requires a further investigation and a discussions about this issue will be given in Section \ref{sec:discussion}.\\
%show that it is crucial to use the parabolic reconstruction and the modified CFL condition to get a second-order accuracy.\\
Computations stop at $t=0.5$ and the piston is  located at  $x=0.75$ for both cases.

%We continue to check the numerical accuracy if the ratio of spatial size and time increment. We choose 
We check the numerical accuracy under the condition that the time increment is chosen to be
\begin{equation}\label{eq:CFL-modified}
\Dt = \text{CFL} \dfr{\Dx^{\frac 32}}{\la_\text{max}},
\end{equation}
where $\la_\text{max}$ is the maximum signal speed. 
The CFL number is taken to be $0.6$.
Since the entropy is uniform for both cases, it is chosen for the comparison.
As shown in Table \ref{tab:constant} for the first case and in Table \ref{tab:accelerated} for the second, the current MBT with the acceleration element achieves desired accuracy of order.

\begin{table*}[!htb]
  \centering
  \caption[small]{Numerical errors and  convergence rates of entropy for the uniform velocity case where $u_c(t)\ev 0.5$. CFL=0.6.}
  \begin{tabular}{|r|l|l|l|l|l|l|l|l|}
    \hline
mesh &      \multicolumn{4}{|c|}{MBT with one-sided Riemann solver }   &      \multicolumn{4}{|c|}{MBT with one-sided GRP solver }     \\\cline{2-9}
  size &\small $L_1$ errors&\small$  L_1 $ orders&\small $  L_\infty $ errors&\small $  L_\infty $ orders&\small $ L_1 $ errors&\small $ L_1 $ orders&\small $ L_\infty $ errors&\small  $ L_\infty $ orders\\\hline
1/40  &  2.49e-5  &            &  4.39e-4  &&  1.54e-5  &         &  8.48e-5  &             \\
1/80  &  4.11e-6  &  2.60   &  1.65e-4  &  1.41 &  1.82e-6  &  3.08   &  6.75e-6       &  3.65 \\
1/160  &  7.34e-7  &  2.48   &  6.09e-5  &  1.44&  2.06e-7  &  3.15   &  7.28e-7     &  3.21  \\
1/320  &  1.44e-7  &  2.35   &  2.29e-5  &  1.41&  2.21e-8  &  3.21   &  7.65e-8     &  3.25    \\
1/640  &3.09e-8  &  2.33   &  8.80e-6  &  1.38 &  2.37e-9 &  3.22    &  7.97e-9  &  3.26        \\
1/1280  &  7.16e-9  &  2.21   &  3.47e-6  &  1.34&  3.03e-10  &  2.97   &  1.01e-9  &  2.98 \\\hline
  \end{tabular}
  \label{tab:constant}
\end{table*}

\begin{table*}[!htb]
  \centering
  \caption[small]{Numerical errors and  convergence rates of entropy for the accelerate case where $u_c(t)=2t$. CFL=0.6.}
  \begin{tabular}{|r|l|l|l|l|l|l|l|l|}
    \hline
mesh &        \multicolumn{4}{|c|}{MBT with one-sided Riemann solver}   &\multicolumn{4}{|c|}{MBT with one-sided GRP solver}     \\\cline{2-9}
size   &\small $L_1$ errors&\small$  L_1 $ orders&\small $  L_\infty $ errors&\small $  L_\infty $ orders&\small $ L_1 $ errors&\small $ L_1 $ orders&\small $ L_\infty $ errors&\small  $ L_\infty $ orders\\\hline1/40      &  4.21e-4  &            &  1.73e-3  &     &  8.77e-5  &         &  1.12e-3  &       \\
1/80      &  1.42e-4  &  1.57  &  1.28e-3  &  0.44&  1.27e-5  &  2.79   &  1.73e-4  &  2.70 \\
1/160    &  5.77e-5  &  1.30   &  9.71e-4  &  0.40&  2.23e-6  &  2.51   &  1.80e-5  &  3.27 \\
1/320    &  2.56e-5  &  1.17   &  7.56e-4  &  0.36 &  4.02e-7  &  2.47   &  3.08e-6  &  2.55 \\
1/640    &  1.18e-5  &  1.12   &  6.03e-4  &  0.33 &  6.32e-8  &  2.67   &  5.97e-7  &  2.37 \\
1/1280  &  5.48e-6  &  1.10   &  4.95e-4  &  0.29 &  8.52e-9  &  2.89   &  1.38e-7  &  2.11 \\\hline  
  \end{tabular}
  \label{tab:accelerated}
\end{table*}

\section{Numerical experiments}\label{sec:numer}

In this section, we will present several typical examples to demonstrate the performance of the current MBT method.  The first two examples are one-dimensional, while the last two are two-dimensional. 
These examples are already benchmark problems in this field, as referred accordingly.    For all the four cases, the CFL number is taken to be $0.6$.

\subsection{Sudden motion of a piston with a rarefaction wave and a shock} 

This benchmark test is taken from  \cite{Shyue-08}.
An infinitely long tube is filled with gas, initially having the  state  $\rho_0=1$, $p_0=5/7$ and $u_0=0$. A piston with the width $2.5$ is initially centered at $18.5$. The piston suddenly  moves to right with a constant speed $u_c=2$, and then a shock forms ahead of the piston and a rarefaction wave behind the piston.
\vspace{0.2cm}

\begin{figure}[!htb]
\centering
 \includegraphics[width=.48\textwidth]{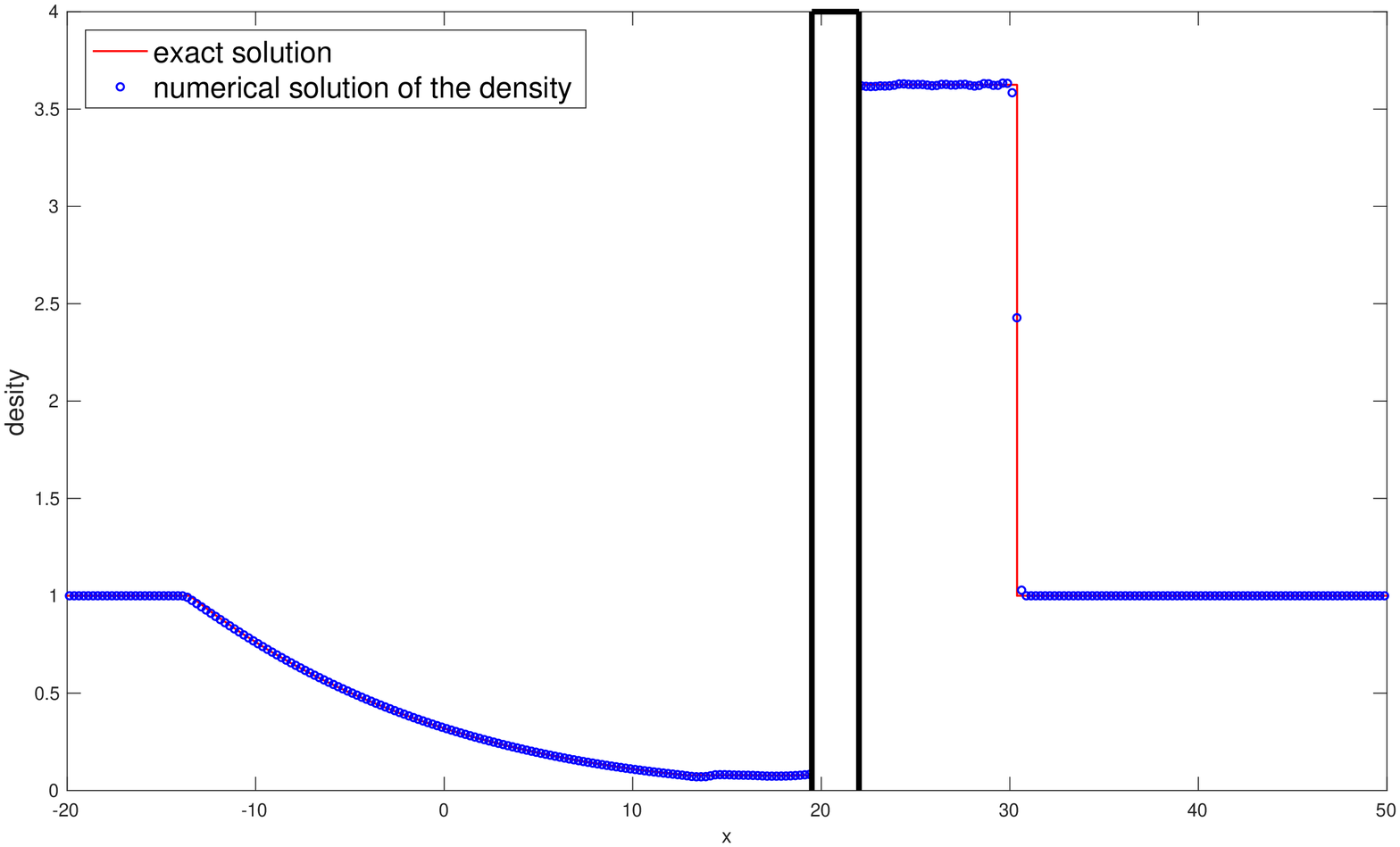}
 \includegraphics[width=.48\textwidth]{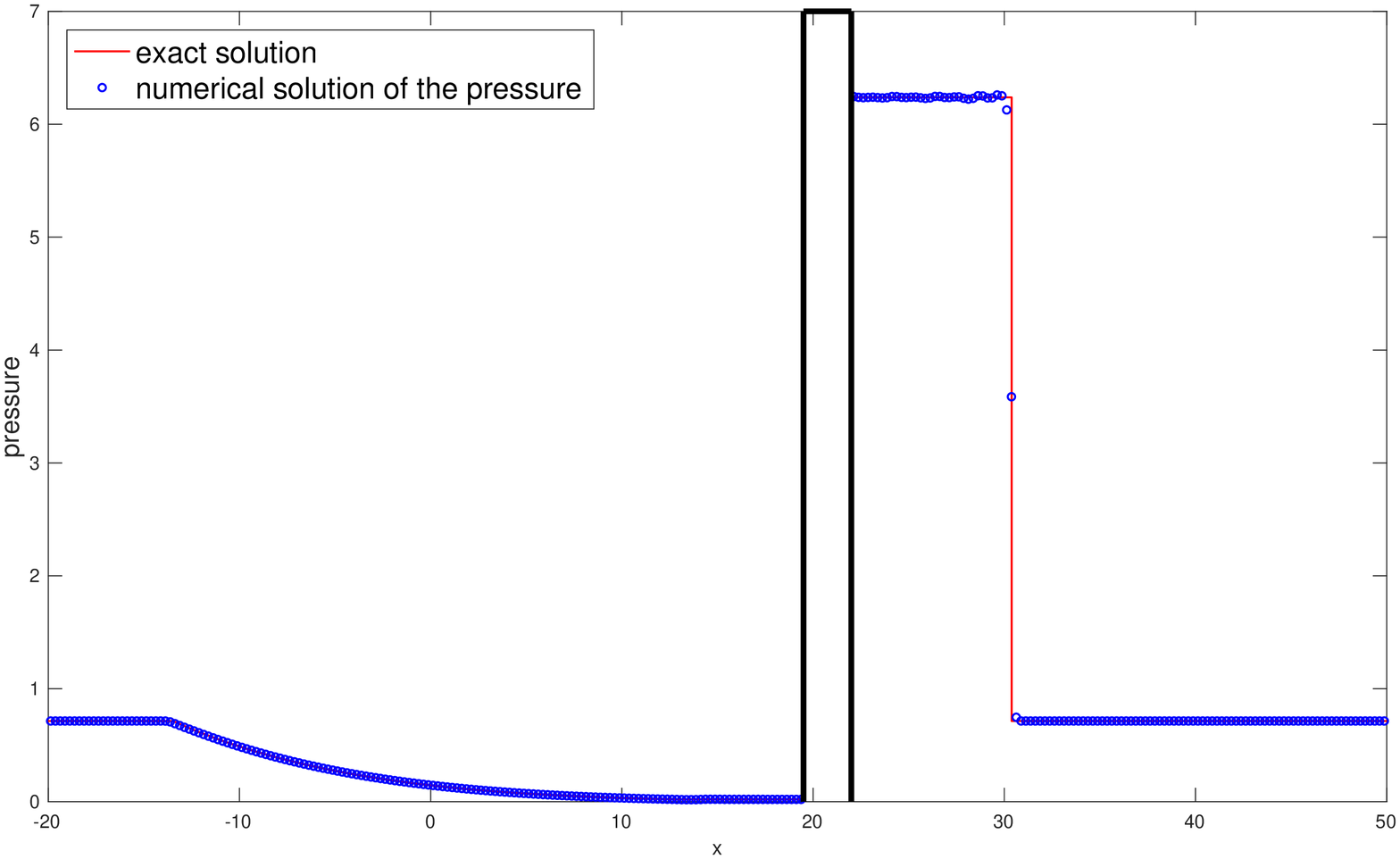}
 \caption[small]{The density (left) and the pressure (right) profiles for  the sudden motion of a piston. }
\label{fig:reimann}
\end{figure}

The computational domain is set to be $[0,70]$, divided into equally distributed cells with the spatial size $\Dx=0.25$.  The numerical results of the density and pressure are shown in Figure \ref{fig:reimann} and both of them match well with the exact solution.\\

\subsection{Sod shock interacting with a piston}   We take this example from \cite{CFSI}. 
A tube occupies the space interval $[0,3]$, filled with gas at rest initially,
\begin{equation*}
  (\rho, u, p)(x,0)=\left\{
\bga{llll}
(1,  ~ 0,  ~ 1),  & x < 1,\\
(0.125,  ~ 0,  ~ 0.1),  & x > 1.
\eda
\right.
\end{equation*}
Both ends of the tube are rigid walls. The computational domain $[0,3]$ is uniformly divided into $600$ cells. This tube is separated by a  piston with the thickness $0.2$, centered initially  at $1.5$.
The motion of the piston   passively depends on  the force exerted from the fluid. 

\begin{figure}[!htb]
\label{fig:spring}
\centering
 \includegraphics[width=.4\textwidth]{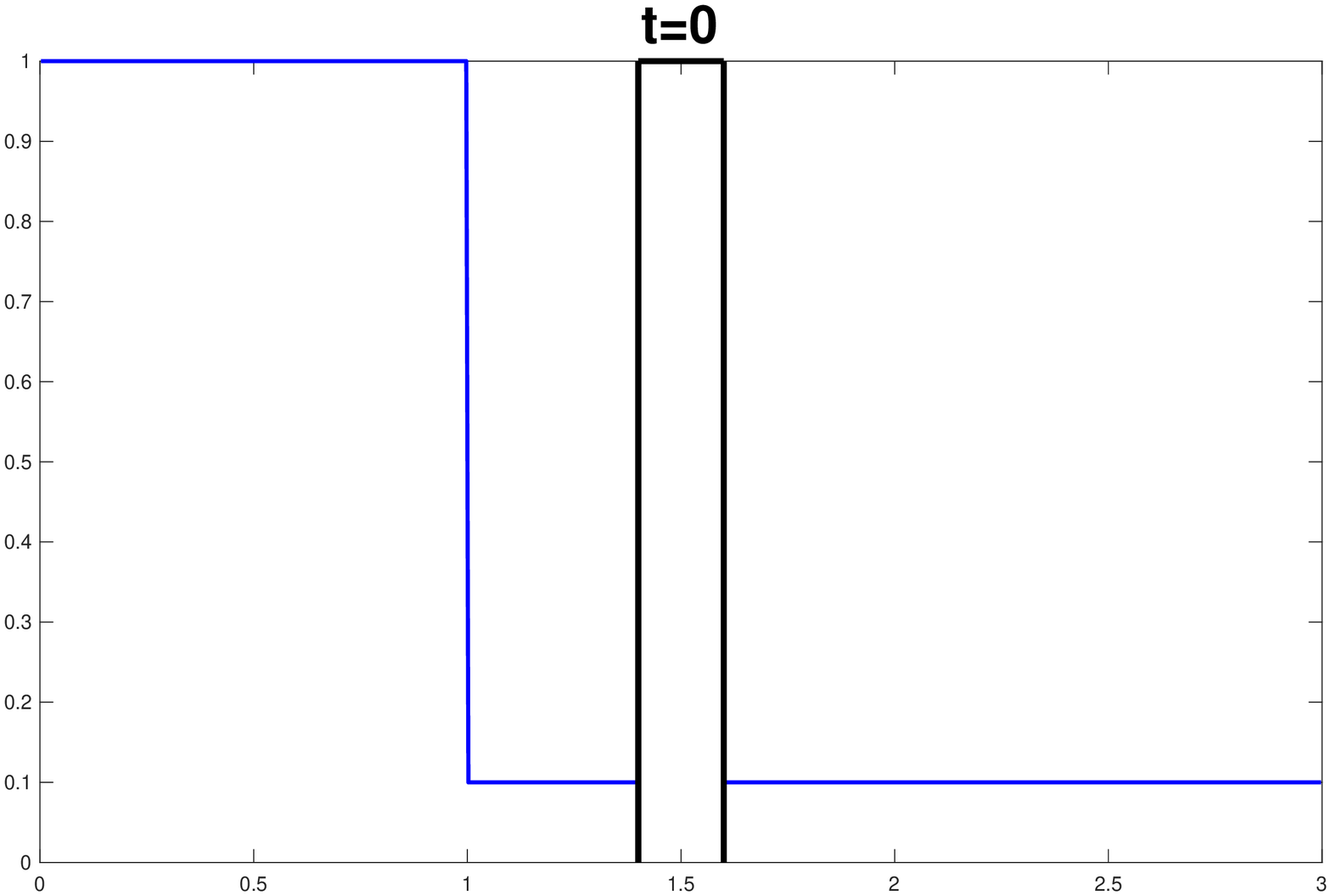}
 \includegraphics[width=.4\textwidth]{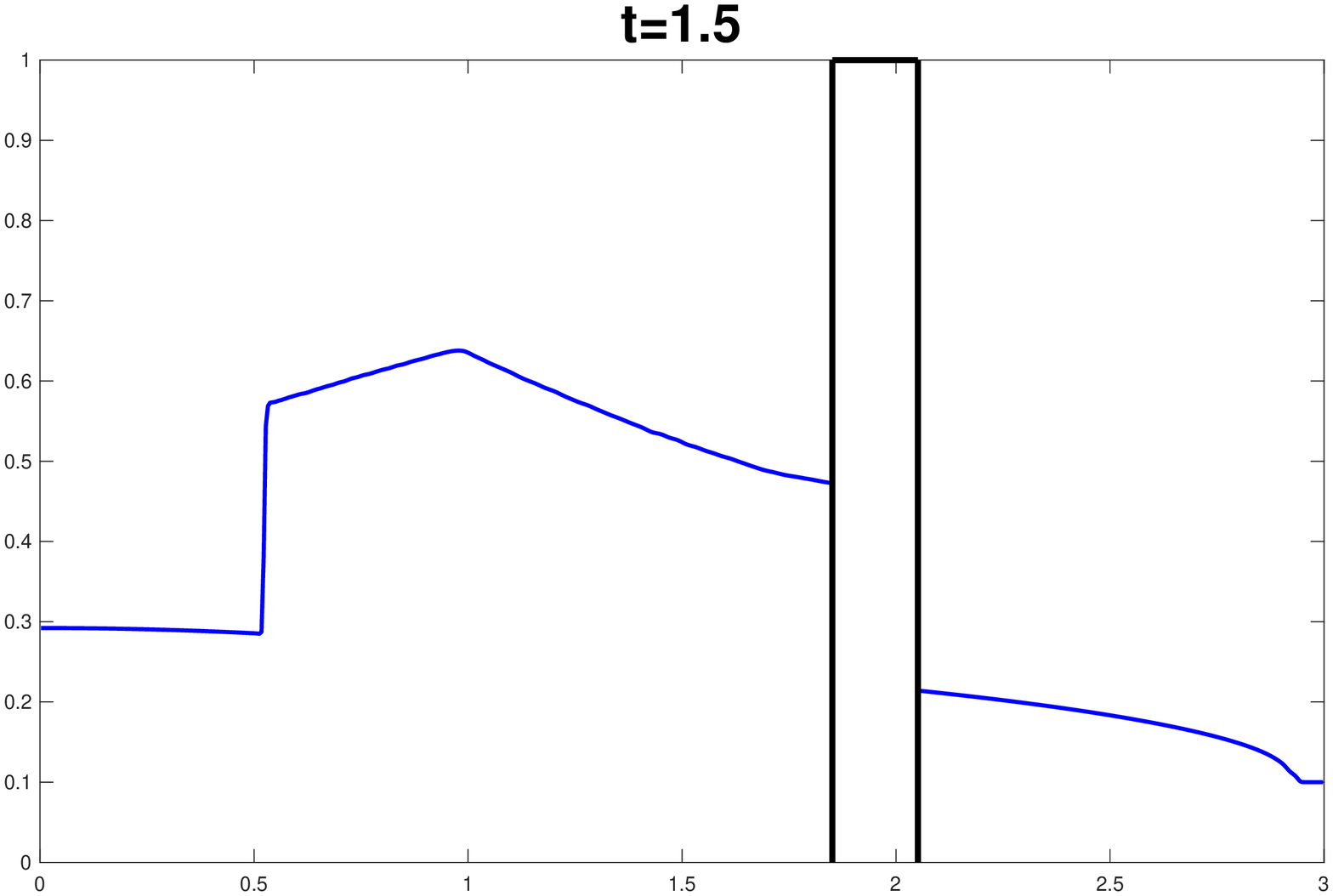}
 \includegraphics[width=.4\textwidth]{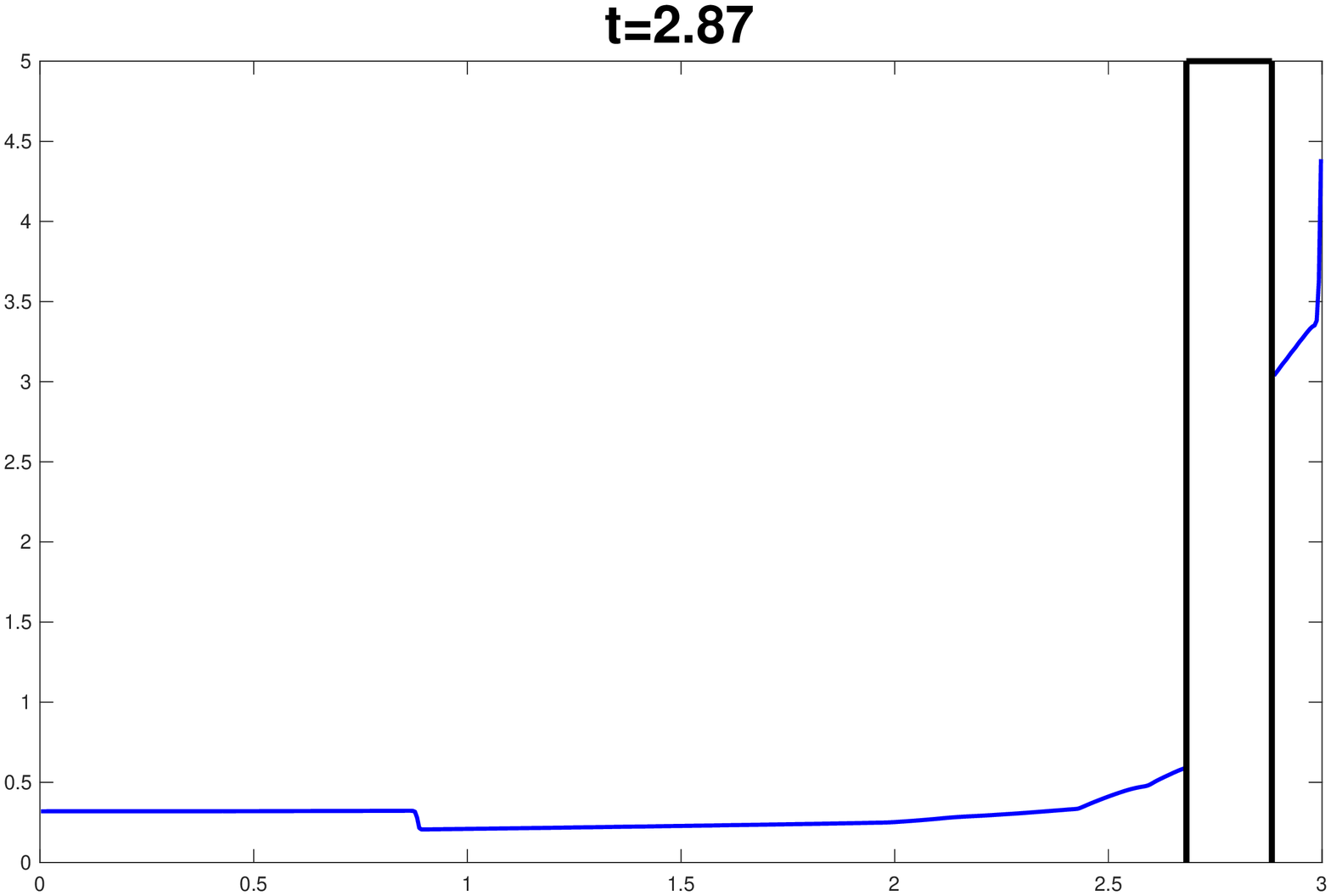}
 \includegraphics[width=.4\textwidth]{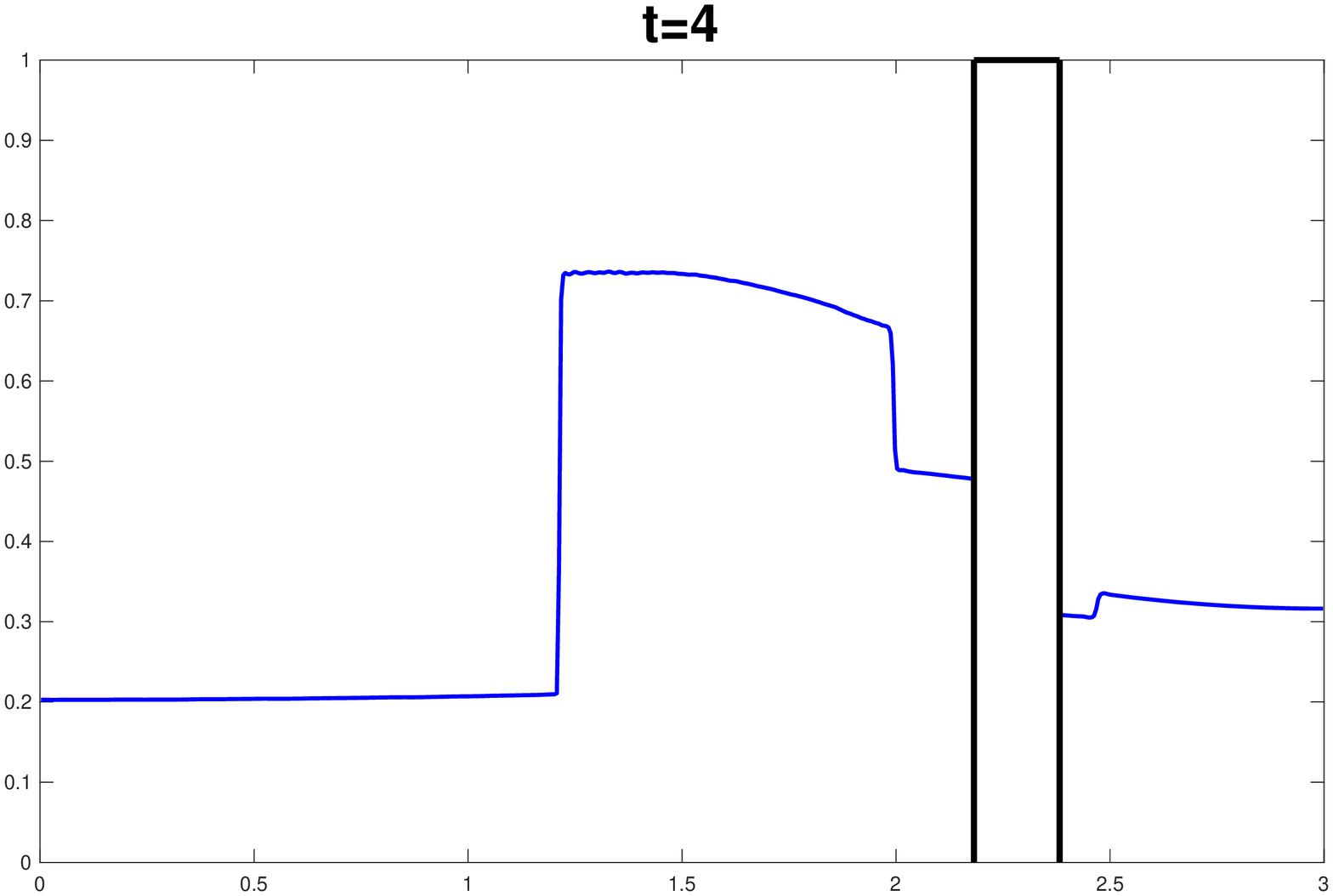}
 \caption[small]{The piston  position and the pressure profile for the Sod shock interacting with a piston at $t=0$, $t=1.5$,  $t=2.87$ and $t=4$.}
\end{figure}

The fluid flow is described as follows:  A shock from the left pushes the piston to the right. Since the right end of the tube is closed, the gas is compressed and the pressure increases as the piston moves right. Finally the piston is bounced back from the right. 
Figure \ref{fig:spring} shows the snapshots of the pressure of the flow field and the piston positions  at different times $t=0$, $t=1.5$,  $t=2.87$,  $t=4$, respectively. The results matches well with those in \cite{CFSI}.\\

\subsection{The kick off of 2-D objects} We consider an oval disk and a rectangular object interacting with a shock, respectively.
A tunnel is $20$ units high, whose top and bottom boundaries are both reflective walls. 
Initially, a shock of Mach number $3$ is posed along $x=8$ and starts to move right. The pre-shock state of the gas is $(\rho_0,u_0,v_0,p_0)=(1,0,0,1)$.

\begin{figure}[!htb]
\centering
 \includegraphics[width=.8\textwidth]{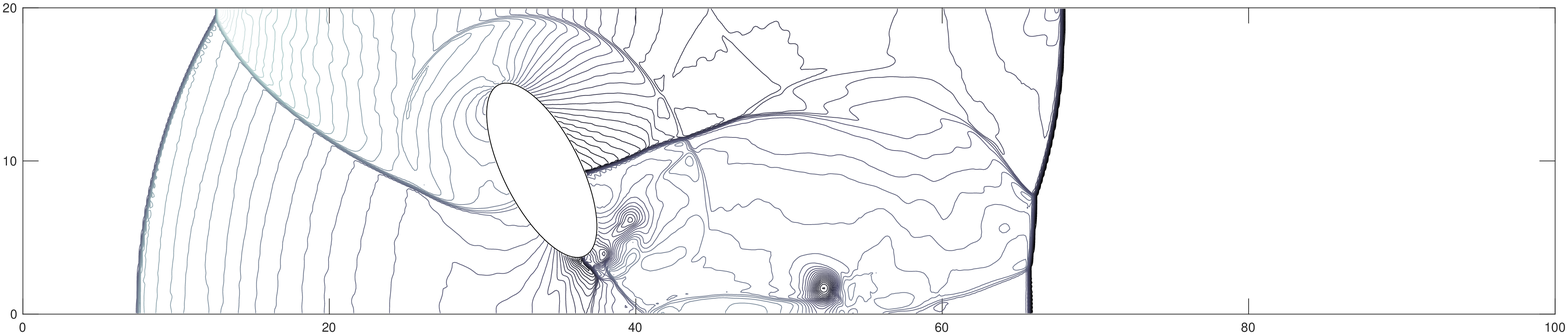}
 \includegraphics[width=.8\textwidth]{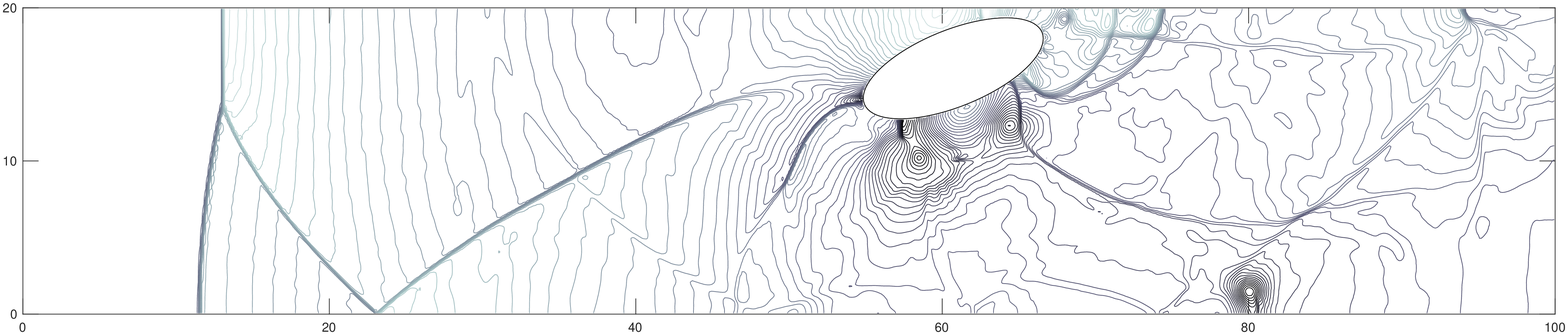}
 \caption[small]{The position of the oval disk and the pressure of the flow field at $t=60$ (upper) and  $t=100$ (lower).}
\label{fig:oval-800x160}
\end{figure}

At first, consider an oval disk of the density $\rho_\Om=10\rho_0$, whose barycenter is  placed at $(15,3)$. The two axis of the oval disk are $a=6.25$ and $b=2.5$ units, respectively. The long axis of the oval disk is placed along the $x$-axis. The mass of the disk is $\mathcal{M}_\Om=\pi a b \rho_\Om$ and the inertia of the disk is $\mathcal{A}_\Om=\frac{\mathcal{M}_\Om}{4}(a^2+b^2)$. 
As the shock moves right, the disk would be kicked off from the ground, and rotates as it moves in the flow field.
The numerical computation is performed in the region $[0,100]\times[0,20]$ divided by a $800\times 160$ grid.
Figure \ref{fig:oval-800x160} shows the position of the disk and the pressure of the flow field at  $t=60$ and $t=100$, respectively. Multiple reflected shocks and vortices formed near the disk can be observed.
The numerical results match those proposed in \cite{Ben-Artzi-01} perfectly with better resolutions.\\

\begin{figure}[!htb]
\centering
 \includegraphics[width=.8\textwidth]{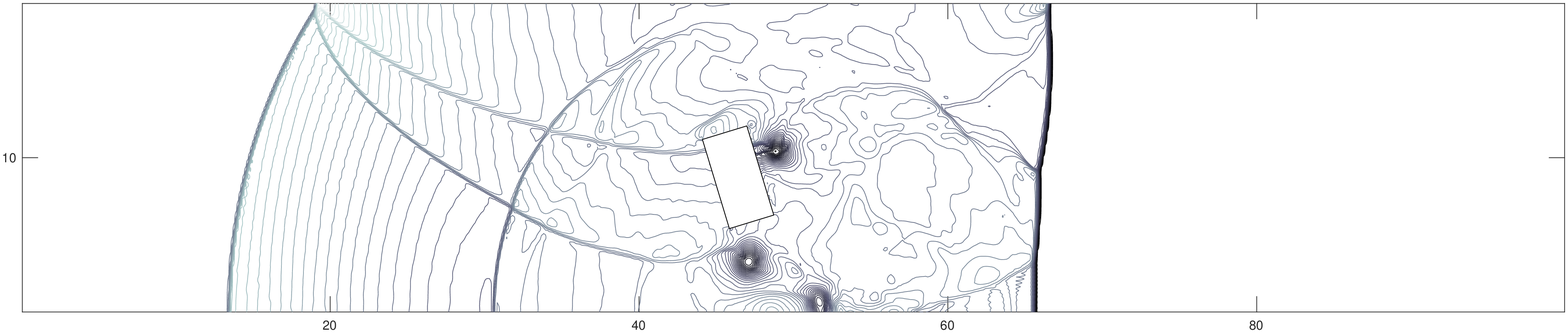}
 \includegraphics[width=.8\textwidth]{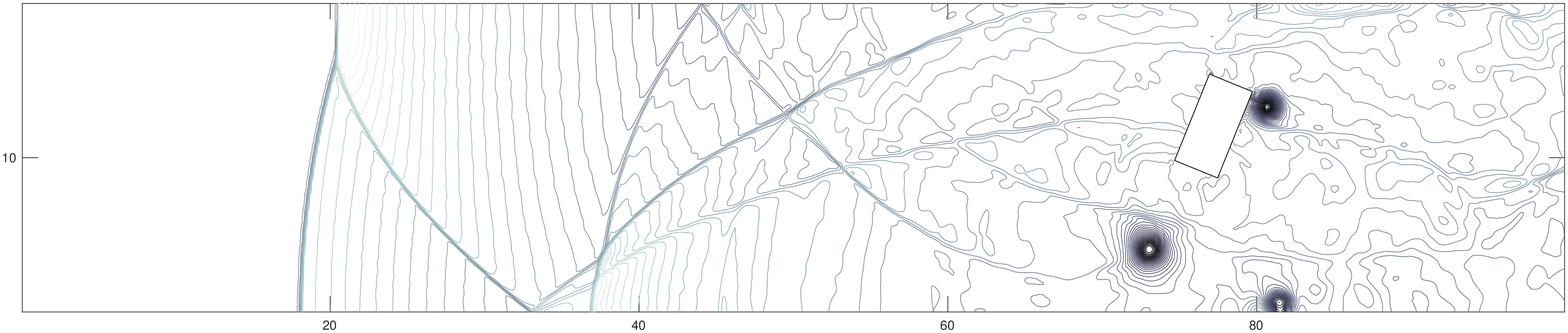}
 \caption[small]{The position of the rectangular object and the pressure of the flow field at $t=60$ (upper) and $t=100$ (lower).}
\label{fig:rect-800x160}
\end{figure}

Now we replace the oval disk by a rectangular object whose barycenter initially locates at $(15,6)$. 
The long  and short sides of the rectangle are $a=6$ and $b=3$ units, respectively.
The long side of the rectangle is posed at an angle of $\pi/4$ with the negative $x$-direction. The mass and the inertia of the rectangle are $\mathcal{M}_\Om=a b \rho_\Om$ and $\mathcal{A}_\Om=\frac{\mathcal{M}_\Om}{3}(a^2+b^2)$, where $\rho_\Om=10\rho_0$.
As the shock moves right, the rectangular object is kicked off. A computation is performed on a $800\times 160$ grid.
The positions of the rectangle and the pressure of the flow field are displayed in Figure \ref{fig:rect-800x160}, at $t=60$ and $t=100$, respectively. Compared with the oval disk, a rectangular object has a larger inertia with respect to its size. Therefore it rotates less, as shown numerically. Besides, the cell-merging procedure is more involved. \\

\subsection{Two cylinders interacting with a shock}  
\begin{figure}[!htb]
\centering
\includegraphics[width=.48\textwidth]{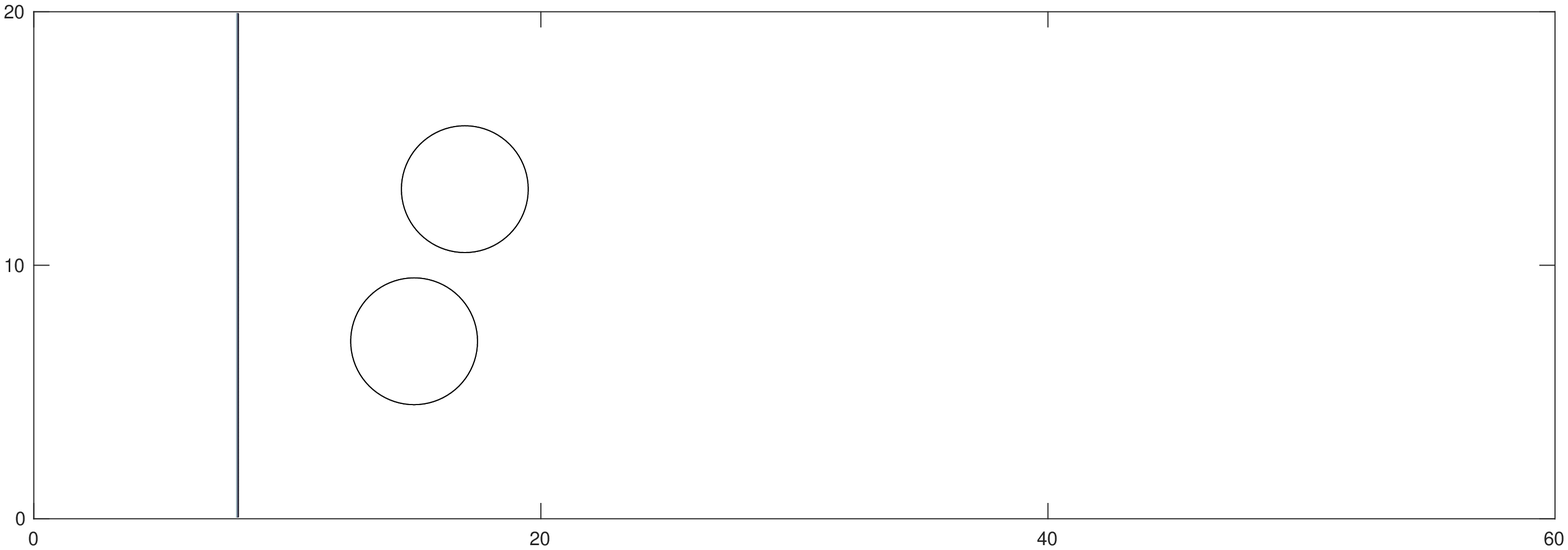}
\includegraphics[width=.48\textwidth]{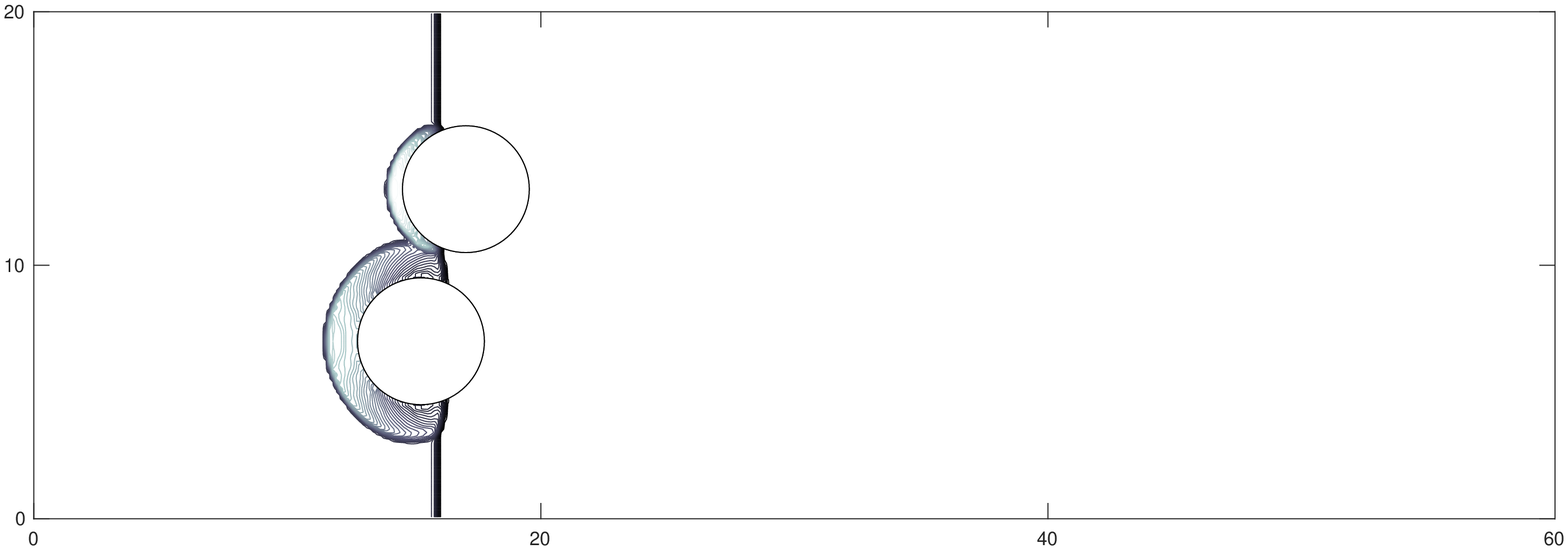}\\
\ \includegraphics[width=.48\textwidth]{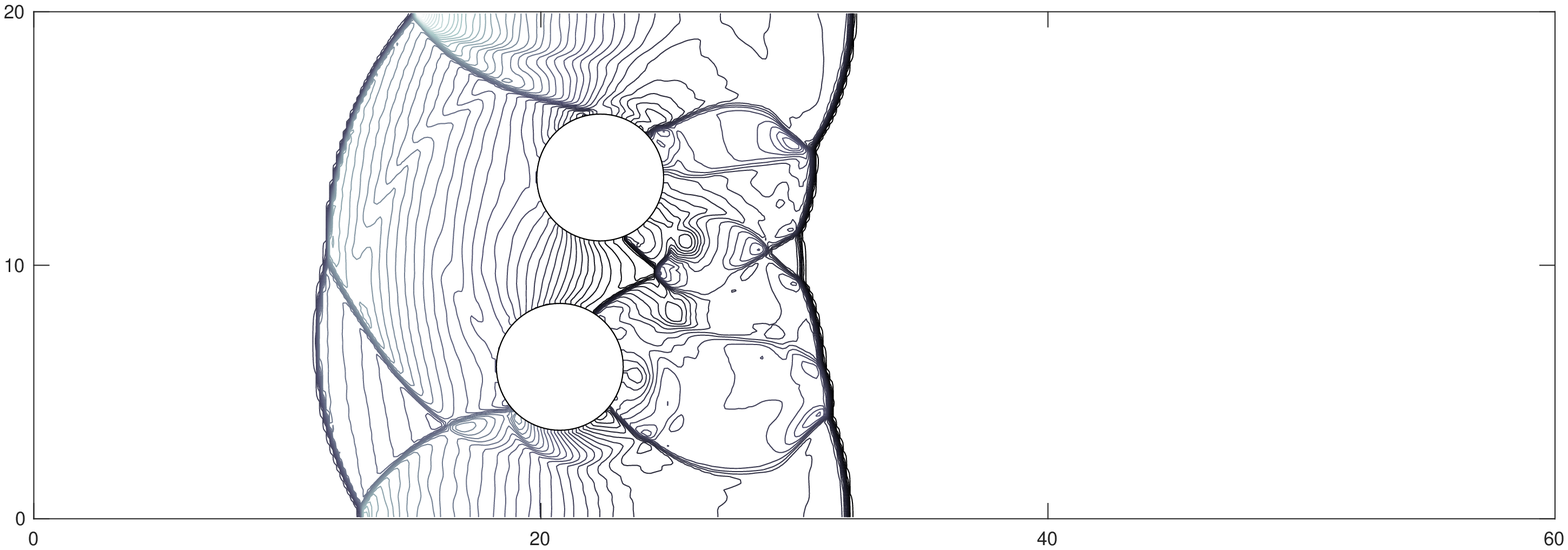}
\includegraphics[width=.48\textwidth]{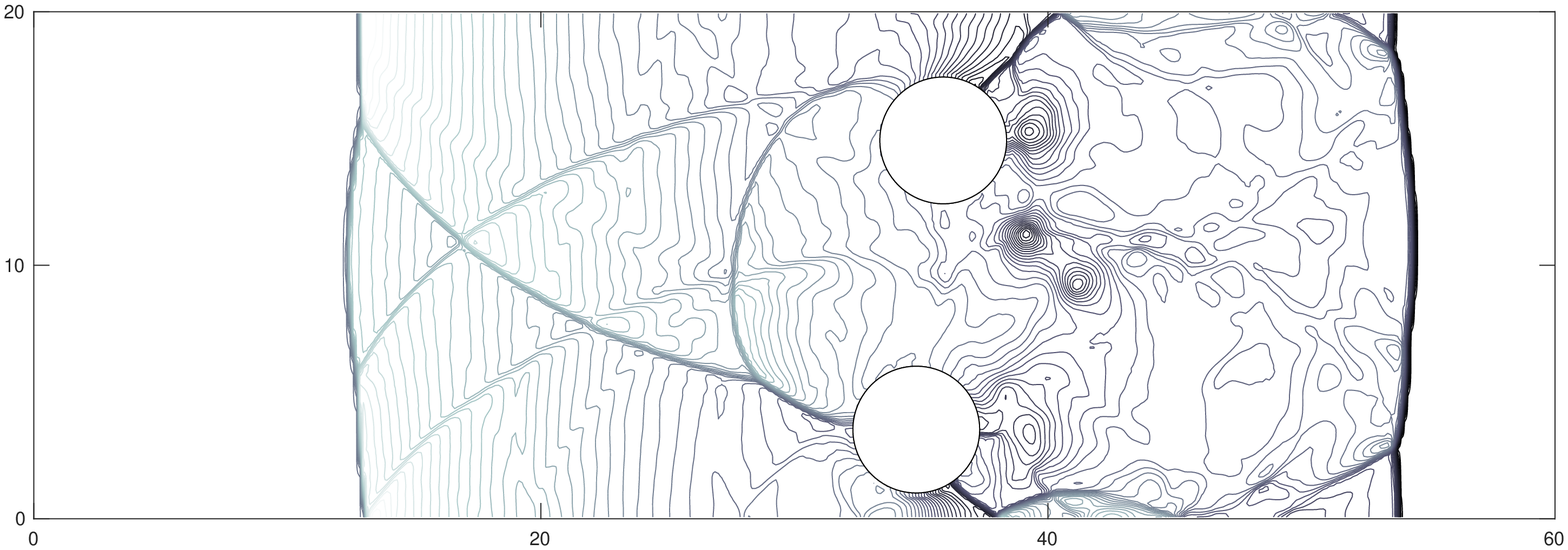}
 \caption[small]{The pressure of the flow field and the positions of two cylinders at $t=0$ (upper left), $t=8$ (upper right), $t=24$ (lower left) and $t=48$ (lower right).}
\label{fig:double-ball}
\end{figure}

We simulate a more complicated problem with a shock interacting with two cylinders. Once again consider a tunnel $20$ units high whose top and bottom boundaries are both reflective walls. A Mach $3$ shock initially locates at $x=8$. The pre-shock fluid state is $(\rho_0,u_0,v_0,p_0)=(1.3,0,0,0.1)$. The two cylinders are positioned with  barycenters at $(15,7)$ and $(19,13)$, respectively. The radius and the density of the cylinders are $r_\Om=2.5$ and $\rho_\Om=10\rho_0$. A computation is carried out in the region $[0,60]\times[0,20]$ with a $450\times 150$ grid.
The pressure of the flow field and positions of two cylinders at $t=0$, $t=8$, $t=24$ and $t=48$ are displayed in Figure \ref{fig:double-ball}.
It can be seen that the two cylinders are pushed away by the higher pressure between them.

\vspace{2mm}
\section{Discussions}\label{sec:discussion} In this paper we observe from the classical piston problem in gas dynamics  that the acceleration is an important element in the description of moving boundaries, and develop a one-sided GRP solver to design a high order moving boundary tracking algorithm, extending  the MBT scheme originally developed in \cite{mbt-falc}. As analyzed,  the newly developed MBT scheme is of second-order, and exhibits the desired performance. 

As  concluding remarks, we would like to make the following discussions. 
\vspace{0.2cm}

\begin{enumerate}
\item[(i)]   It is a natural choice to use the  pair $(u_c,\frac{du_c}{dt})$, the velocity and the acceleration,  to describe the motion of moving boundaries.  In literature, this observation is not  directly integrated into the design of high order schemes, possibly due to the absence of the one-sided GRP solver.  Of course, this solver can be equivalently derived as in the context of gas kinetic scheme \cite{Xu-2001} or approximated, e.g., in \cite{mbt-ilw}, with applications in other framework such as  the adaptive grid method \cite{Olim-1993,mbt-adaptive} and the over-lapping grid method \cite{overlap-NASA,overlap-Fedkiw}.
\vspace{0.2cm}

\item[(ii)]  This method can be also extended to higher order versions, such as the fourth order version using the two-stage fourth order framework in \cite{Du-Li-1, Li-2019}. This is left for the further study. 

\vspace{0.2cm}

\item[(iii)] The present paper just studies the moving boundary tracking of solid bodies. Careful check of the methodology stated in Section 2 indicates that this method can be extended to track various types of moving boundaries, e.g., in the context of multi-material flows and  viscous flows etc. The key ingredient is how to evaluate the force exerted on the moving surface to compute out the acceleration.

\end{enumerate}

\vspace{2mm}

\vspace{2mm}
\bibliographystyle{siam}
\bibliography{reference}

\end{document}

%% file: abbr-notation.tex
\def\ptext{\text{p}}

\def\Dt{\De t}
\def\Dx{\De x}
\def\Dy{\De y}

\def\ba{\mathbf{a}}

\def\bn{\mathbf{n}}

\def\br{\mathbf{r}}

\def\bu{\mathbf{u}}

\def\bx{\mathbf{x}}

\def\bF{\mathbf{F}}
\def\bG{\mathbf{G}}

\def\bT{\mathbf{T}}

\def\bW{\mathbf{W}}
\def\bX{\mathbf{X}}

\def\n{\noindent}
\def\pt{\partial}

\def\Re{\mathbb{R}}

\def\f#1#2{\frac {#1}{#2}}
\def\f32{\frac 32}

\def\d{\displaystyle}

\def\bga{\begin{array}}
\def\eda{\end{array}}

\def\De{\Delta}

\def\ev{\equiv}

\def\Gm{\Gamma}
\def\gm{\gamma}

\def\th{\theta}
\def\nb{\nabla}

\def\la{\lambda}

\def\al{\alpha}

\def\rw{\rightarrow}
\def\iy{\infty}
\def\Om{\Omega}
\def\om{\omega}

\def\d{\displaystyle}
\def\dfr#1#2{\displaystyle{\frac{#1}{#2}}}